\documentclass[sigconf]{acmart}
\AtBeginDocument{%
  }
    
\copyrightyear{2026}
\acmYear{2026}
\setcopyright{cc}
\setcctype{by}
\acmConference[CIKM '26]{Proceedings of the 35th ACM International Conference on Information and Knowledge Management}{November 07--11, 2026}{Rome, Italy}
\acmBooktitle{Proceedings of the 35th ACM International Conference on Information and Knowledge Management (CIKM '26), November 07--11, 2026, Rome, Italy}
\acmDOI{10.1145/3799682.3841126}
\acmISBN{979-8-4007-2539-5/2026/11}

\usepackage{colortbl}
\usepackage{subcaption}
\usepackage{multirow}
\usepackage{enumitem}

\newcommand{\modelname}{PRISM}

\begin{document}

\title{From Overlooked to Explored: Recovering Item Relations via Mixture of Perspectives for Sequential Recommendation
}
\author{Junyoung Kim}
\email{junyoungkim@postech.ac.kr}
\affiliation{%
  \institution{Pohang University of \\ Science and Technology}
  \city{Pohang}
  \country{Republic of Korea}
}
\author{Wonbin Kweon}
\email{wonbinkweon@skku.edu}
\affiliation{%
  \institution{Sungkyunkwan University}
  \city{Suwon}
  \country{Republic of Korea}
}
\author{Woojoo Kim}
\email{kimuj0103@postech.ac.kr}
\affiliation{%
  \institution{Pohang University of \\ Science and Technology}
  \city{Pohang}
  \country{Republic of Korea}
}
\author{Jaehyung Lim}
\email{jaehyunglim@postech.ac.kr}
\affiliation{%
  \institution{Pohang University of \\ Science and Technology}
  \city{Pohang}
  \country{Republic of Korea}
}
\author{Dongha Kim}
\email{dhkim0317@postech.ac.kr}
\affiliation{%
  \institution{Pohang University of \\ Science and Technology}
  \city{Pohang}
  \country{Republic of Korea}
}
\author{Hwanjo Yu}
\email{hwanjoyu@postech.ac.kr}
\authornote{Corresponding author.}
\affiliation{%
  \institution{Pohang University of \\ Science and Technology}
  \city{Pohang}
  \country{Republic of Korea}
}

\renewcommand{\shortauthors}{Junyoung Kim et al.}
\begin{abstract}
\label{abstract}
Capturing user preference from a user's interaction sequence is the central challenge of Sequential Recommendation (SR).
This preference intuitively emerges from inter-item relations: each item transition reflects a preference embedded in the relations between items, making the faithful capture of these relations essential for accurate recommendation.
For this reason, self-attention is dominant in sequential recommendation for its ability to compute pairwise item interactions, yet our empirical analysis reveals that it consistently suffers from \textit{similarity bias} across various types of transformer-based SR models: dot-product attention scores disproportionately favor similar items, systematically overlooking heterogeneous relations with meaningful preference signals and directly limiting recommendation performance.
To address this, we propose \textbf{PRISM} (\textbf{P}erspective-based \textbf{R}elational \textbf{I}nsight \textbf{S}ynthesis \textbf{M}odule), a module that re-examines item relations from multiple perspectives.
PRISM employs $K$ \textit{Perspective Lenses} to calibrate attention from distinct viewpoints, combining an \textit{Affinity View} that refines homogeneous relations and a \textit{Contrast View} that exposes heterogeneous ones suppressed by similarity bias, enabling the model to capture the full spectrum of user preferences.
Extensive experiments on seven real-world benchmarks demonstrate that PRISM consistently outperforms state-of-the-art baselines.
Our code is available at \textcolor{blue}{https://github.com/327aem/PRISM/}.
\end{abstract}




\maketitle
\section{Introduction}
\begin{figure}[t]
    \centering
    \includegraphics[width=\columnwidth]{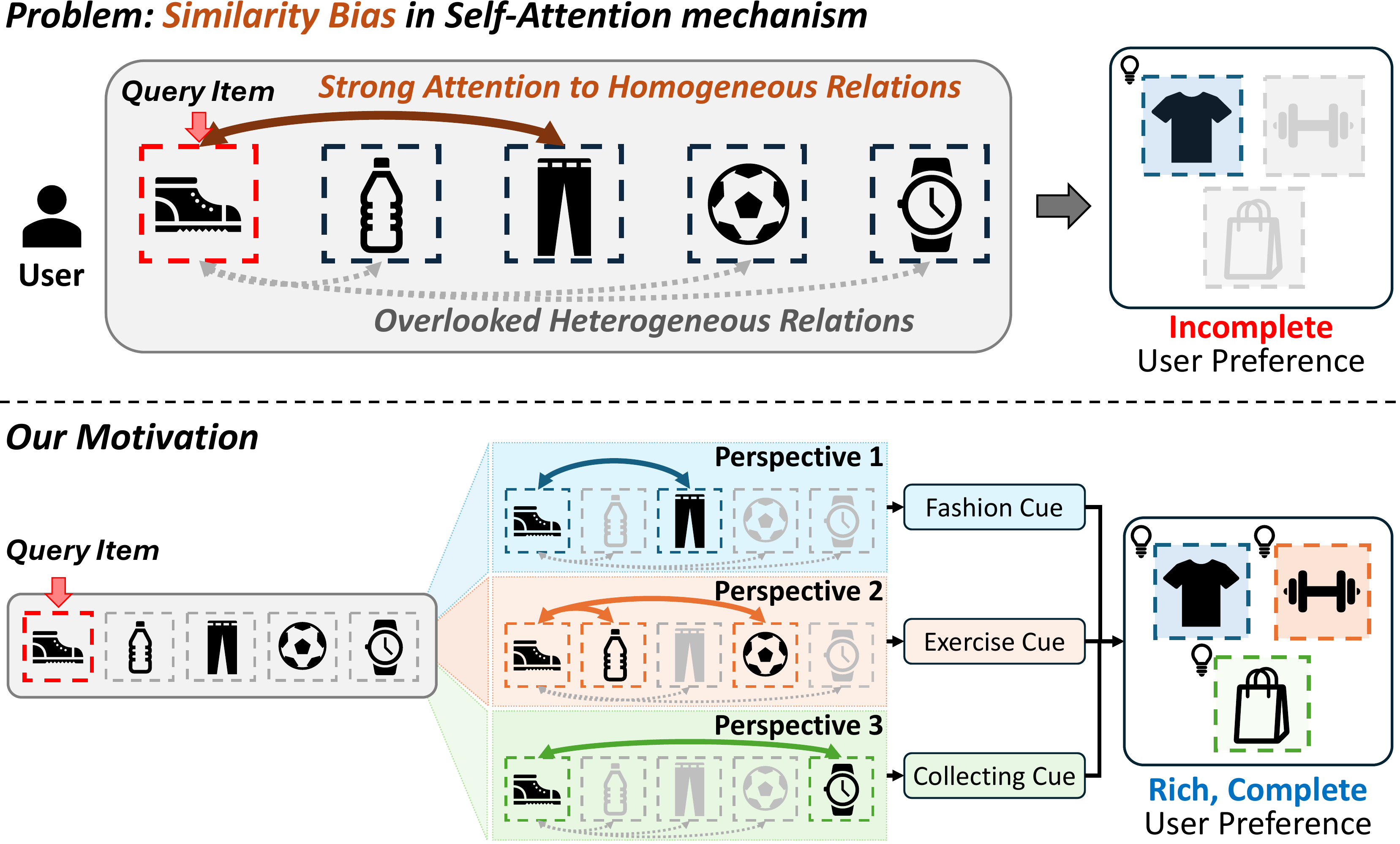}
    \caption{Motivation of PRISM. (Top) Standard self-attention favors homogeneous relations, capturing only a partial preference. (Bottom) Examining the sequence from multiple perspectives, each surfacing a distinct cue (e.g., Fashion, Exercise, Collecting), can combine them into a richer preference.}
    \label{fig:motivation}
\end{figure}
Sequential Recommendation (SR) aims to predict a user's next item from their interaction history.
Given diverse and dynamic consumption patterns, the central challenge in SR is to faithfully model the user's \textit{preference} driving these interactions from the sequence.

In practice, this preference surfaces through inter-item relations: when a user transitions from one item to the next, that choice reflects a specific preference embedded in the relation between the two items, and which preference each item carries depends on its surrounding context.
Capturing the user's preference, therefore, calls for a closer look at these inter-item relations.
For this reason, transformer-based models~\cite{sasrec, bert4rec, trm1, trm2, god, trm3, trm4} have become dominant in SR.
By computing pairwise interactions across all items in the sequence, self-attention weighs how much each item relates to every other, making it naturally suited for capturing inter-item relations and achieving strong recommendation performance.

Despite this success, prior work~\cite{actsr} has reported that self-attention often produces erroneous attention distributions.
We argue that this stems from how attention is computed: attention scores are dot-product similarities between queries and keys, so each item's representation is dominated by items already similar to it, while dissimilar ones contribute little.
We call this tendency \textit{similarity bias} that overlooks heterogeneous relations.
As described in Fig.~\ref{fig:motivation} (top), given a query item of running shoes, attention inherently concentrates on the relation to the pants, similar to the shoes, while relations to the relatively dissimilar water bottle, soccer ball, and watch are suppressed.
As a result, only the \textit{fashion} cue is captured, and additional meaningful cues such as \textit{exercise} or \textit{collecting} are missed, leaving the model unable to capture the diverse preferences embedded in the user's sequence.

We empirically verify this in Sec.~\ref{sec:analysis} across four transformer-based baseline categories: vanilla self-attention~\cite{sasrec}, attention calibration for correcting attention scores~\cite{actsr}, intent-based modeling for capturing user intent~\cite{iclrec,icsrec}, and mixture-of-experts for multi-faceted preference modeling~\cite{fame}.
Our analyses reveal that items important for target prediction are systematically overlooked within self-attention.
As shown in Sec.~\ref{sec:overlooked}, the correlation between each item's attention score and its actual contribution to target prediction is consistently low, and in some cases even negative; this misalignment persists across all four categories.
Furthermore, in Sec.~\ref{sec:targeted_recovery}, when we artificially boost attention on the overlooked items, the predicted probability of the target item increases.
These overlooked relations carry meaningful preference signals, and recovering them calls for a mechanism that complements standard attention.

To this end, we propose \textbf{\modelname} (\textbf{P}erspective-based \textbf{R}elational \textbf{I}nsight \textbf{S}ynthesis \textbf{M}odule), a module placed between transformer layers.
Analogous to an optical prism that decomposes a single beam of light into a spectrum of colors, PRISM employs $K$ \textit{Perspective Lenses} to calibrate attention at each layer from multiple semantic perspectives, recovering the relations that the preceding transformer layer overlooked.
Each lens operates in one of two complementary views: the \textit{Affinity View} refines homogeneous relations already captured by attention, while the \textit{Contrast View} surfaces heterogeneous relations suppressed by similarity bias.
In Fig.~\ref{fig:motivation} (bottom), one \textit{Perspective Lens} connects the shoes with the pants for a \textit{fashion} cue, another with the water bottle and the soccer ball for an \textit{exercise} cue, and another with the watch for a \textit{collecting} cue.
Combining these cues recovers the full spectrum of user preference that similarity bias would otherwise suppress.
Through successive layers, the user's preference emerges more clearly.
Extensive experiments on diverse benchmarks demonstrate that PRISM consistently outperforms existing baselines and proves robust across various real-world scenarios.
Our contributions are as follows:
\begin{itemize}[leftmargin=*]
    \item We identify \textit{similarity bias} as a cause of overlooked relational preference signals.
    \item We propose \textbf{PRISM}, whose $K$ \textit{Perspective Lenses} refine homogeneous relations through the \textit{Affinity View} and recover suppressed heterogeneous ones through the \textit{Contrast View}.
    \item Experiments on diverse benchmarks show that PRISM outperforms baselines and remains robust across real-world scenarios.
\end{itemize}

\begin{figure*}[t]
    \centering
    \includegraphics[width=\linewidth]{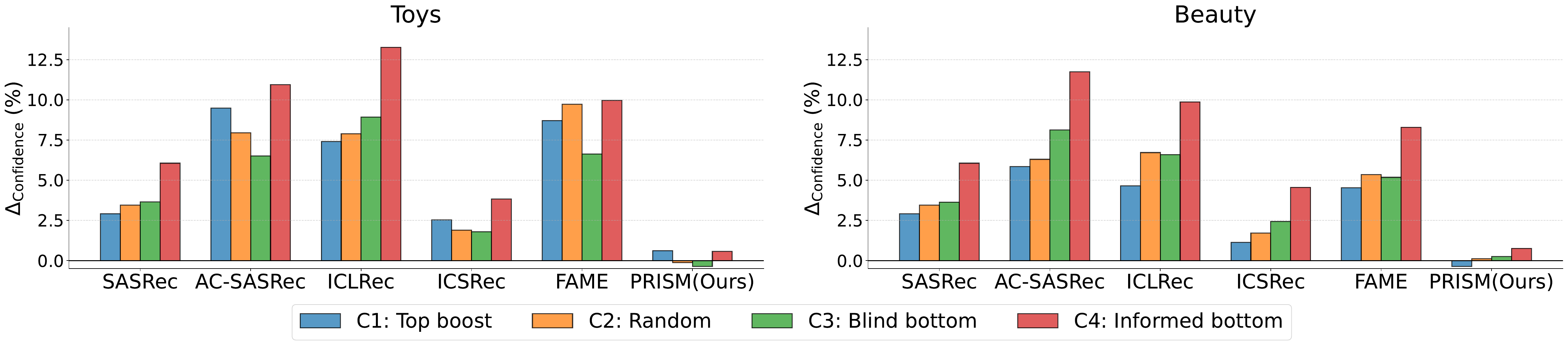}
    \caption{Analysis of counterfactual attention intervention. $\Delta_{\text{confidence}}$ (\%) when boosting attention at positions in the top 20\% by attention (C1), a random 20\% (C2), the bottom 20\% (C3), and the bottom 20\% that are also top-50\% by causal importance (C4).}
    \label{fig:recovery}
\end{figure*}

\section{Related Works}
\subparagraph{\textbf{Transformer-based SR.}}
Recommendation aims to model user--item relationships across diverse settings, including federated~\cite{fed1, fed2, fed3, f3crec} and continual~\cite{conti1, conti2, conti3}, while recent work also leverages LLMs~\cite{llm1,llm2,llm3,tracer, llm2rec, vlm2rec}.
In particular, Sequential Recommendation (SR) predicts the next item by modeling relations within a user's interaction sequence.
Transformer-based models like SASRec~\cite{sasrec} and BERT4Rec~\cite{bert4rec} dominate SR by leveraging self-attention to capture inter-item relations~\cite{caser,fmlp,factormodel,gru4rec,gruplus}.
AC-SASRec~\cite{actsr} further calibrates attention scores to correct erroneous attention distributions.
However, such calibration relies on a learnable matrix without explicit relational guidance.
Our PRISM calibrates attention from multiple perspectives, using inter-item relations as explicit guidance to recover the heterogeneous relations.
\subparagraph{\textbf{Intent-based SR.}}
ICLRec~\cite{iclrec}, IOCRec~\cite{iocrec}, ICSRec~\cite{icsrec}, and ELCRec~\cite{elcrec} capture user intent by assigning each sequence a single global intent representation learned via contrastive learning.
Modeling preference at the sequence level, however, suppresses item-level relational signals that drive individual transitions; furthermore, as they typically build on self-attention, they inherit similarity bias.
PRISM complements sequence-level modeling by operating at the item level: $K$ Perspective Lenses analyze each item's relational context from distinct semantic viewpoints, enabling heterogeneous preference signals to surface within the sequence.
\subparagraph{\textbf{Mixture-of-Experts for Recommendation.}}
MoE architectures~\cite{noisygating,moe2,moe3,moe4,moduleformer} have been widely adopted in recommendation, including multi-task~\cite{mmoe,rel2,snr}, multi-modal~\cite{mm1,mm2,mm3}, and SR~\cite{fame,famousrec,starrec}.
Sequence-level MoE models (e.g., STAR-Rec~\cite{starrec}, FamouSRec~\cite{famousrec}) route sequences to different experts, while FAME~\cite{fame} treats each attention head as a distinct item facet and applies MoE within each head.
Although FAME leverages multiple heads to capture multi-faceted item properties, each head still computes dot-product attention, so similarity bias persists within every facet.
PRISM addresses this through its dual-view design: the \textit{Affinity View} refines homogeneous relations and the \textit{Contrast View} explicitly recovers the heterogeneous relations that similarity bias suppresses.
\section{Motivating Analysis: \\Overlooked Relations in Self-Attention}
\label{sec:analysis}

We empirically test whether self-attention overlooks relations that are important for predicting the target item.

\begin{figure}
    \centering
    \includegraphics[width=\columnwidth]{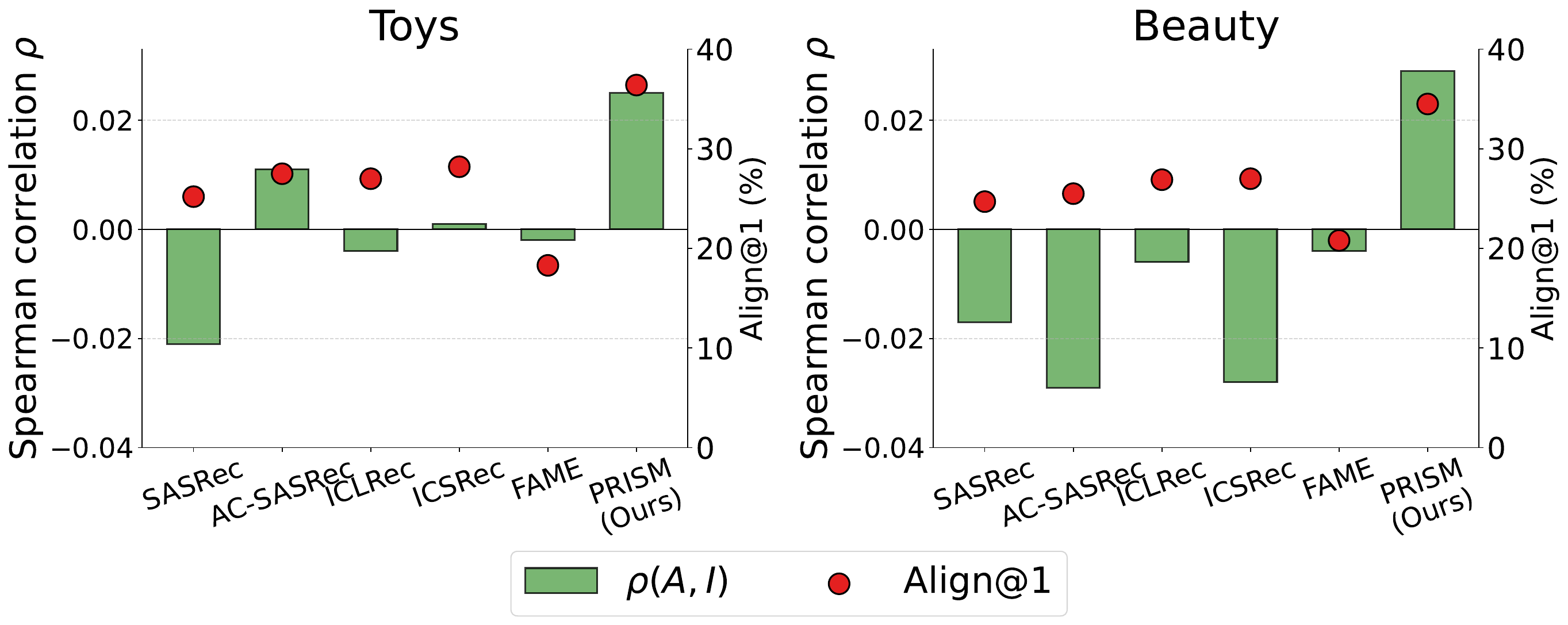}
    \caption{Attention-Importance alignment across models.}
    \label{fig:alignment}
\end{figure}

\subsection{Problem Setup}

SR aims to predict a user's next item from their historical interaction sequence.
Let $\mathcal{U}$ and $\mathcal{V}$ denote the sets of users and items.
For a user $u \in \mathcal{U}$, we denote the sequence of user $u$ as $\mathcal{S}^{u} = [v_1, v_2, \dots, v_{|\mathcal{S}^{u}|}]$, where $v_t \in \mathcal{V}$ is the item interacted with at step $t$.

\begin{equation}
\hat{v}_{|\mathcal{S}^{u}|+1} = \arg\max_{v \in \mathcal{V}} P(v \mid \mathcal{S}^{u}).
\end{equation}

\noindent Each item is mapped to a $d$-dimensional embedding via the embedding table $\mathbf{E} \in \mathbb{R}^{|\mathcal{V}| \times d}$.
A stack of $L$ transformer layers~\cite{transformer} produces representations $\mathbf{H}^l \in \mathbb{R}^{T \times d}$ at each layer $l \in \{1, \dots, L\}$ with max sequence length $T$ via
\begin{equation}
\begin{gathered}
\mathbf{A}^l = \frac{\mathbf{Q}^l (\mathbf{K}^l)^{\top}}{\sqrt{d_k}} \in \mathbb{R}^{T \times T}, \\
\mathbf{H}^l = \text{FeedForward}(\text{Softmax}(\text{CM}(\mathbf{A}^l))\, \mathbf{V}^l),
\end{gathered}
\label{eq:attn}
\end{equation}
where the query, key, and value matrices are $\mathbf{Q}^l = \mathbf{H}^{l-1} \mathbf{W}_q^l$, $\mathbf{K}^l = \mathbf{H}^{l-1} \mathbf{W}_k^l$, and $\mathbf{V}^l = \mathbf{H}^{l-1} \mathbf{W}_v^l$, respectively, $d_k$ denotes the dimensionality of the key vectors, and $\text{CM}(\cdot)$ denotes the causal masking.
The last hidden state of the last transformer layer, $\mathbf{h}_u := \mathbf{H}^L_{T,\,:} \in \mathbb{R}^d$, serves as the sequence representation.

\subsection{Attention Overlooks Important Items}
\label{sec:overlooked}

\subsubsection{\textbf{Measuring Item Importance.}}
To identify which items most contribute to predicting the target, we measure each position's \textit{causal importance}.
For a sequence $\mathcal{S}^u$ with ground-truth next item $y$, the causal importance of position $j$ is the drop in target probability (i.e., model confidence toward target item) when that item is masked:
\begin{equation}
I_j(\mathcal{S}^u) = P(y \mid \mathcal{S}^u) - P(y \mid \text{mask}(\mathcal{S}^u, j)),
\label{eq:causal_importance}
\end{equation}
where $\text{mask}(\mathcal{S}^u, j)$ masks the item of $\mathcal{S}^u$ at position $j$.
A large $I_j$ means that item $j$ causally contributes to predicting the target, regardless of how much attention it receives.

\subsubsection{\textbf{Setup.}}
We extract the last-layer, head-averaged attention logit $A_j = \frac{1}{\#\text{heads}}\sum_{\text{head}} A^{\text{head}}_{T,j}$ ($T$: the last query position used for prediction) and compute $I_j$ for 1{,}000 test sequences on the Amazon Toys and Beauty datasets.
We evaluate five transformer-based SR models: vanilla self-attention (SASRec), attention calibrated (AC-SASRec, which directly corrects attention scores), intent-based models (ICLRec, ICSRec), and MoE within attention heads (FAME).
Within each sequence, we compute $\rho(A, I)$, the Spearman rank correlation between attention and causal importance, and Align@1, the fraction of sequences where the top-attended position matches the most causally important one (i.e., $\arg\max_j A_j = \arg\max_j I_j$).

\subsubsection{\textbf{Results.}}
To assess whether attention aligns with causal importance, we report $\rho(A, I)$ and Align@1 across all baselines (Fig.~\ref{fig:alignment}), and observe the following.

\subparagraph{\textbf{Finding 1: Attention misses causally important items.}}
In SASRec, which uses only dot-product self-attention, $\rho(A, I)$ is negative on both Toys and Beauty, and Align@1 is similarly low, indicating that dot-product attention favors items similar to the query rather than items the model actually relies on for prediction.

\subparagraph{\textbf{Finding 2: Existing designs do not close this gap.}}
AC-SASRec recalibrates attention scores, yet $\rho(A, I)$ worsens on Beauty compared to SASRec. FAME adds MoE within attention heads, yet Align@1 drops below SASRec. ICLRec and ICSRec cluster user preferences, yet the misalignment persists. 
None of these designs revises how attention selects items, and so all fail to close the gap. 
We next ask whether recovering overlooked items improves prediction.

\begin{figure*}[t]
    \centering                                                                                                    
    \includegraphics[width=0.95\textwidth]{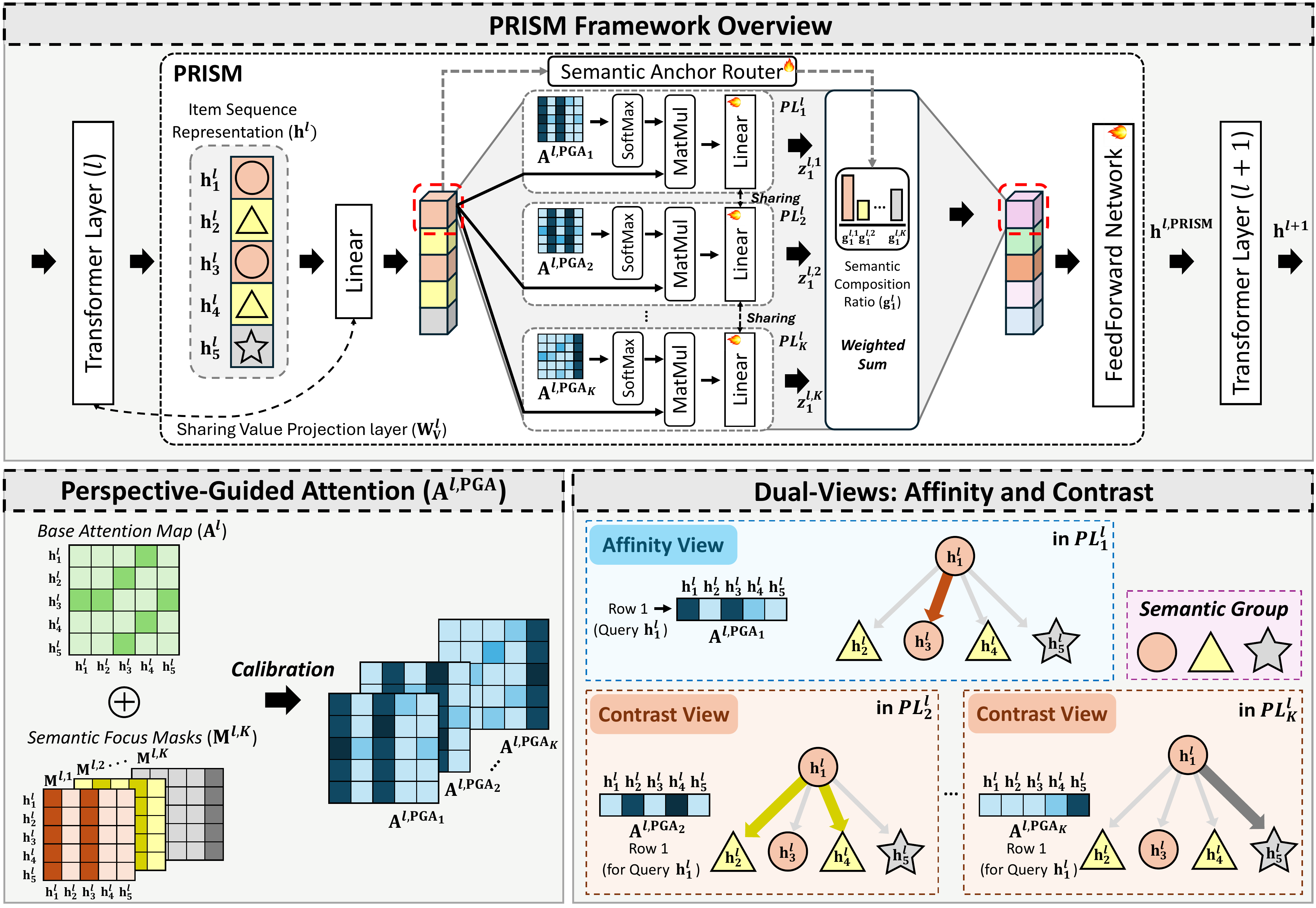}
    \caption{PRISM framework: PRISM performs item-wise multi-perspective analysis followed by relational insight synthesis. All learnable parameters are shared across lenses for efficiency, while the outputs differ solely by relational perspective.}
    \label{fig:arch1}
\end{figure*}

\subsection{Recovering Overlooked Items \\Improves Prediction}
\label{sec:targeted_recovery}

\subsubsection{\textbf{Defining Intervention Conditions.}}
To examine which types of items benefit most from increased attention, we define four intervention conditions:
\begin{itemize}[nosep,leftmargin=*]
    \item \textbf{C1 (Top boost):} boost the top-20\% positions by attention.
    \item \textbf{C2 (Random):} boost a random 20\% of positions.
    \item \textbf{C3 (Blind bottom):} boost the bottom-20\% by attention, recovering items overlooked by attention.
    \item \textbf{C4 (Informed bottom):} boost only the bottom-20\% positions that are also in the top-50\% by causal importance $I_j$, selectively recovering items that are overlooked yet causally important.
\end{itemize}

\subsubsection{\textbf{Setup.}}
For each test sequence, we modify the last-layer attention logits $\mathbf{A}^L$ at inference.
Under each condition, each $j \in \mathcal{P}$ (the set of selected positions) is boosted by the mean attention logits over $\mathcal{P}$: $\mathbf{A}^L_{T,j} \leftarrow \mathbf{A}^L_{T,j} + \beta_{\text{boost}} \cdot \frac{1}{|\mathcal{P}|}\sum_{j' \in \mathcal{P}} \mathbf{A}^L_{T,j'}$, with boosting strength $\beta_{\text{boost}} = 5.0$.
We then measure the change in the model's confidence for the target item $\Delta_{\text{confidence}}(\%) = 100 \times (P_{\text{modified}} - P_{\text{original}}) / P_{\text{original}}$.

\subsubsection{\textbf{Results.}}
To test whether recovering overlooked items benefits prediction, we report $\Delta_{\text{confidence}}$ under each intervention condition (Fig.~\ref{fig:recovery}), and observe the following.

\subparagraph{\textbf{Finding 1: All modifications improve prediction.}}
All baselines show large improvement of $\Delta_{\text{confidence}}$ values across all conditions.
Even random perturbation (C2) helps substantially, indicating that baseline attention poorly reflects which items the model actually relies on, so almost any redistribution improves prediction.

\subparagraph{\textbf{Finding 2: Overlooked items contain useful signals.}}
On both datasets, each baseline shows $C3 > 0$: boosting \emph{any} low-attention item improves prediction, confirming that the overlooked positions contain genuinely useful relational signals.

\subparagraph{\textbf{Finding 3: Useful signals span diverse relational facets.}}
C4 consistently outperforms C3: not all overlooked relations contribute equally to target prediction, and useful signals are distributed across diverse relational facets rather than concentrated in a single group.

\subsection{Summary}
\label{sec:analysis_summary}

Together, these analyses reveal that self-attention systematically overlooks inter-item relations that carry meaningful evidence for target prediction, and that these overlooked relations are heterogeneous rather than confined to a single type.
This motivates a mechanism that examines item relations from multiple perspectives, enabling the model to capture user preferences that a single uniform view of the sequence would miss.

\section{Methodology}

To address these limitations, we propose \textbf{\modelname} (\textbf{P}erspective-based \textbf{R}elational \textbf{I}nsight \textbf{S}ynthesis \textbf{M}odule), a module placed between transformer layers that recovers inter-item relations overlooked by a single attention pattern, with $L{-}1$ PRISM modules in an $L$-layer transformer. \modelname\ consists of two parts:
\begin{itemize}[nosep,leftmargin=*]
    \item \textbf{($\S$\ref{sec:architecture}) Model Architecture.} To re-examine the sequence from multiple viewpoints, we first identify each item's semantic composition over $K$ groups ($\S$\ref{sec:sar}). $K$ Perspective Lenses then re-examine the attention from distinct viewpoints, recovering overlooked relations ($\S$\ref{sec:mopl}), and the outputs are synthesized into a unified representation ($\S$\ref{sec:ris}). The resulting representations are integrated into the transformer stack for final prediction ($\S$\ref{sec:prediction}).
    \item \textbf{($\S$\ref{sec:training}) Training Objectives.} We introduce $\mathcal{L}_{\text{SPCL}}$ to complement the item-level analysis with sequence-level preference ($\S$\ref{sec:spcl}), and $\mathcal{L}_{\text{CCL}}$ to ensure cross-lens agreement without collapsing their diversity ($\S$\ref{sec:ccl}).
\end{itemize}
The overall framework is shown in Fig.~\ref{fig:arch1}.

\subsection{Model Architecture}
\label{sec:architecture}
\subsubsection{\textbf{Semantic Anchor Router}}
\label{sec:sar}
To quantitatively measure and classify the complex semantic characteristics of each item's representation, we introduce the \textit{Semantic Anchor Router}.
First, we define the number of semantic groups $K$ and compute the routing logits $\mathbf{r}_i^l \in \mathbb{R}^K$ for item $i$'s representation $\mathbf{h}_i^l$ using a noisy gating mechanism~\cite{noisygating} to encourage exploration in training:
\begin{equation}
    \mathbf{r}_i^l = \mathbf{h}_i^l \mathbf{W}_\mathbf{g}^l + \eta_i^l \odot \text{Softplus}(\mathbf{h}_i^l \mathbf{W}_{\text{noise}}^l), \quad \eta_i^l \sim \mathcal{N}(\mathbf{0}, \mathbf{I}_K),
    \label{eq:logits}
\end{equation}
where $\mathbf{W}_g^l, \mathbf{W}_{\text{noise}}^l \in \mathbb{R}^{d \times K}$ are learnable matrices, $\eta_i^l \in \mathbb{R}^K$ is a Gaussian noise vector, $\mathbf{I}_K \in \mathbb{R}^{K \times K}$ is the identity matrix, and $\odot$ denotes element-wise multiplication. At inference, the noise term is disabled.
Then, we obtain the \textit{Semantic Composition Ratio} ($\mathbf{g}_i^l \in \mathbb{R}^K$):
\begin{equation}
    \mathbf{g}_i^l = \text{Softmax}(\mathbf{r}_i^l).
    \label{eq:gate}
\end{equation}
Each value $\mathbf{g}_i^{l,k}$ represents the degree to which item $i$ belongs to semantic group $k$ ($1 \leq k \leq K$).
Based on this ratio, we identify each item's dominant semantic group, termed the \textit{Primary Semantic Anchor (PSA)}, by selecting the index with the maximum value:
\begin{equation}
    \text{PSA}_i^l = \underset{k \in \{1,...,K\}}{\text{argmax}}(\mathbf{g}_i^{l,k}).
\end{equation}

\subsubsection{\textbf{Mixture of Perspective Lenses}}
\label{sec:mopl}
To address the overlooked relations identified in Sec.~\ref{sec:analysis}, we propose Perspective Lenses ($PL^l_k$), specialized for semantic group $k$, that calibrate the attention logits map $\mathbf{A}^l$ from the previous transformer layer through this group's perspective, adaptively strengthening both homogeneous and heterogeneous item relations.

\subparagraph{\textbf{Semantic Focus Mask.}}
First, each $PL^l_k$ constructs a binary \textit{Semantic Focus Mask}, $\mathbf{M}^{l,k}\in\{0, 1\}^{T \times T}$.
This mask is determined solely by the PSA$_j^l$ at each key position $j$, and is shared across all queries:
\begin{equation}
\mathbf{M}_{ij}^{l,k} = \mathbb{I}(\text{PSA}_{j}^{l} = k),
\end{equation}
where $\mathbb{I}(\cdot)$ is the indicator function.
The resulting mask sets columns for group-$k$ keys to 1, so every query $i$ evaluates the sequence through the same group-$k$ filter in $PL_k^l$.

\subparagraph{\textbf{Relational Signal Boosting.}}
Next, we calculate a \textit{Relational Boost Signal} $\mathbf{S}^{l,k} \in \mathbb R^{T \times T}$ to quantify the group-specific importance for each lens $k$:
\begin{equation}
\label{eq:rsb}
\mathbf{S}_{ij}^{l,k} = \left( \frac{1}{T} \sum_{j'=1}^{T} \mathbf{A}_{ij'}^{l} \cdot \mathbf{M}_{ij'}^{l,k} \right) \cdot \mathbf{M}_{ij}^{l,k}
\end{equation}
For each query item $i$, $\mathbf{S}_{i}^{l,k}$ captures the average attention logits toward all \textit{key} items anchored to the semantic group $k$.

\subparagraph{\textbf{Perspective Guided Attention.}}
We then fuse this signal into the previous attention logits $\mathbf{A}^l$ using a learnable scalar $\beta^l$:
\begin{equation}
\label{eq:pga}
\mathbf{A}^{l,\text{PGA}_{k}} = \mathbf{A}^{l} + \exp(\beta^l \cdot \mathbf{S}^{l,k})
\end{equation}
This yields the final \textit{Perspective Guided Attention (PGA)} logits map for $PL^l_k$, denoted as $\mathbf{A}^{l,\text{PGA}_{k}}$, containing the recalibrated scores for all query-key item relation pairs (Fig.~\ref{fig:arch1}, lower-left).
Here, $\exp(\cdot)$ ensures that the added boost signal is always non-negative.
Consequently, the final representation $\mathbf{Z}^{l,k} \in \mathbb{R}^{T \times d}$ produced by $PL^l_k$ is computed as:
\begin{equation}
    \mathbf{Z}^{l,k} = \text{Linear}^{l}\!\left( \text{Softmax}(\mathbf{A}^{l,\text{PGA}_{k}})\, {\mathbf{V}'}^{l} \right),
    \quad {\mathbf{V}'}^{l} = \mathbf{H}^{l}\mathbf{W}_v^{l},
    \label{eq:z}
\end{equation}
where ${\mathbf{V}'}^{l} \in \mathbb{R}^{T \times d}$, and $\mathbf{z}_i^{l,k} \in \mathbb{R}^d$, the $i$-th row of $\mathbf{Z}^{l,k}$, is the lens output for item $i$.
Importantly, PRISM shares all learnable parameters ($\beta^l$ in Eq.~\ref{eq:pga} and Linear layer in Eq.~\ref{eq:z}) across lenses.
This design guarantees efficiency while ensuring that distinct lens outputs reflect solely the influence of the $k$-th perspective assigned to each $PL^l_k$, via $\mathbf{M}_{ij}^{l,k}$.

\subparagraph{\textbf{How the Dual Views Emerge.}}
This perspective-guided calibration dynamically yields two complementary relational views (Fig.~\ref{fig:arch1} lower-right), depending on the alignment between the query item $i$'s semantic anchor $\text{PSA}_i^l$ and the lens's semantic group $k$.

When query item $i$ belongs to the semantic group of lens $PL_k^l$ (i.e., $\text{PSA}_i^l = k$), the lens operates in the \textbf{Affinity View}.
This lens strengthens homogeneous relations among items in the same semantic group, refining their shared semantics into a more precise form.
For instance, an \textit{Electronics} lens strengthens the relation between a \textit{laptop} and a \textit{monitor}, refining their connection into the more specific concept of \textit{building a workstation}.

Conversely, when query item $i$ falls outside the lens's semantic group $k$ ($\text{PSA}_i^l \neq k$), the lens operates in the \textbf{Contrast View}.
This mode attends to cross-group items otherwise suppressed by \textit{similarity bias}, revealing latent preferences across semantic groups.
For example, a \textit{Travel} lens connects a \textit{laptop} (Electronics) with a \textit{backpack} (Travel) despite their different semantic groups, surfacing the novel concept of a \textit{digital nomad} from item relations that might otherwise be neglected.

Since each item has a unique $\text{PSA}_i^l$, every item is guaranteed to be analyzed under exactly one Affinity View and $K{-}1$ Contrast Views.
Together, the two views jointly cover both within-group and cross-group item relations.

\subsubsection{\textbf{Relational Insight Synthesis}}
\label{sec:ris}
To synthesize the relational insights derived from all $K$ Perspective Lenses, PRISM aggregates the diverse representations $\mathbf{z}_i^{l,k}$ into a single unified representation $\mathbf{h}_i^{l,\text{PRISM}}$.
This is accomplished via a weighted sum using the entries of the \textit{Semantic Composition Ratio} $\mathbf{g}_i^l$ (Eq.~\ref{eq:gate}) as weights, preserving the item's composite semantic membership.
\begin{equation}
\label{eq:ris}
\mathbf{h}_i^{l,\text{PRISM}} = \text{FeedForward}^{l}\!\left(\sum_{k=1}^{K} \mathbf{g}_i^{l,k} \cdot \mathbf{z}_i^{l,k}\right)
\end{equation}
The resulting vector is then processed by a Feed-Forward Network before entering the next transformer layer.

\subsubsection{\textbf{Integration and Prediction}}
\label{sec:prediction}
The representations produced by PRISM, $\mathbf{H}^{l,\text{PRISM}} = [\mathbf{h}_1^{l,\text{PRISM}}, \dots, \mathbf{h}_T^{l,\text{PRISM}}]$, are passed as input to the $(l{+}1)$-th transformer layer.
The sequence representation $\mathbf{h}_u \in \mathbb{R}^d$ from the last transformer layer and the item embedding matrix $\mathbf{E}$ are used to compute the prediction over all items:
\begin{equation}
\hat{\mathbf{y}}_u = \text{Softmax}(\mathbf{h}_u \mathbf{E}^\top) \in \mathbb{R}^{|\mathcal{V}|}.
\end{equation}

\subsection{Training Objective}
\label{sec:training}

\subsubsection{\textbf{Sequence-level Preference Contrastive Loss}}
\label{sec:spcl}

PRISM captures preferences at the item-relation level, but preferences also manifest at the broader sequence level.
To incorporate this, we propose $\mathcal{L}_{\text{SPCL}}$, a contrastive loss that aligns sequences at the shared sequence-level preference without collapsing the distinct relational patterns each lens captures.

\subparagraph{\textbf{Same-Target Augmentation.}}
Following~\cite{duorec, icsrec}, we adopt same-target augmentation to effectively capture sequence-level preference: for each training sequence, we retrieve another sequence whose target item is identical, forming a positive pair.
However, directly applying contrastive learning on these pairs causes \textit{over-alignment}: the loss aligns sequences not only at the shared goal but also at perspective-specific relations, collapsing the diversity that distinguishes users with different relational patterns.

\subparagraph{\textbf{Stochastic Perspective Masking.}}
To prevent over-alignment, we introduce \textit{Stochastic Perspective Masking} (SPM), which preserves the primary semantic signal while stochastically masking non-primary lenses, regularizing the contrastive alignment.
At each PRISM layer $l$, for item $i$ we generate a binary mask $\mathbf{m}_{i}^{l} \in \{0, 1\}^K$:
\begin{equation}
\label{eq:spm}
\mathbf{m}_{i}^{l,k} \sim
    \begin{cases}
        \text{Bernoulli}(1 - \rho), & \text{if } k \neq \text{PSA}_{i}^{l} \\
        1, & \text{if } k = \text{PSA}_{i}^{l}
    \end{cases}
\end{equation}
where $\rho$ is the perspective dropout rate.
This mask is applied to the routing logits $\mathbf{r}_{i}^{l}$ (Eq.~\ref{eq:logits}) via a log-mask operation:
\begin{equation}
    \tilde{\mathbf{r}}_{i}^{l} = \mathbf{r}_{i}^{l} + \log(\mathbf{m}_{i}^{l} + \epsilon),
\end{equation}
which drives masked entries to near-zero probability after softmax, ensuring that only the sequence-level preference is aligned while each user's perspective-specific relations remain distinct.

\subparagraph{\textbf{Contrastive Learning.}}
The masked gate $\tilde{\mathbf{g}}_i^{l} = \text{Softmax}(\tilde{\mathbf{r}}_i^{l})$ flows through Eq.~\ref{eq:ris} and the remaining layers, yielding the anchor and positive sequence representations $\tilde{\mathbf{h}}_a, \tilde{\mathbf{h}}_p$ at the last transformer layer (as in Sec.~\ref{sec:prediction}).
$\mathcal{L}_{\text{SPCL}}$ is then defined as:
\begin{equation}
    \mathcal{L}_{\text{SPCL}} = -\log \frac{\exp\left(\text{sim}(\tilde{\mathbf{h}}_a, \tilde{\mathbf{h}}_p)/\tau\right)}{\sum_{b=1}^{B} \exp\left(\text{sim}(\tilde{\mathbf{h}}_a, \tilde{\mathbf{h}}_b)/\tau\right)},
\end{equation}
where $\text{sim}(\cdot,\cdot)$ denotes dot-product similarity, $\tau$ is a temperature, and $B$ is the batch size.

\subsubsection{\textbf{Collaborative Consistency Loss}}
\label{sec:ccl}

Because each Perspective Lens receives the same item as input to analyze with distinct perspectives, their outputs should share a consistent item-space interpretation while preserving each lens's unique perspective; to enforce this, we propose $\mathcal{L}_{\text{CCL}}$.
Specifically, for each lens output $\mathbf{z}_i^{l,k}$ of item $i$ (Eq.~\ref{eq:z}), we compute a similarity distribution $p_i^{l,k}$:
\begin{equation}
p_i^{l,k} = \text{Softmax}(\mathbf{z}_i^{l,k} \mathbf{E}_{\text{sample}}^\top) \in \mathbb{R}^N,
\end{equation}
where $\mathbf{E}_{\text{sample}} \in \mathbb{R}^{N \times d}$ contains sampled item embeddings and $N = |\mathbf{E}_{\text{sample}}|$.

We then minimize the Symmetric KL divergence $\text{SKL}(p, q) := \tfrac{1}{2}[\text{KL}(p\|q) + \text{KL}(q\|p)]$ between lens pairs and average across the $L{-}1$ PRISM modules:
\begin{equation}
\mathcal{L}_{\text{CCL}} = \frac{1}{L-1}\sum_{l=1}^{L-1}\sum_{i=1}^{T} \sum_{1 \leq k < k' \leq K} w_{i}^{l,k,k'} \cdot \text{SKL}(p_i^{l,k}, p_i^{l,k'}).
\end{equation}

Since not all lenses contribute equally to item $i$'s representation, we weight each pair by their semantic composition ratios $\mathbf{g}_i^{l,k}$ (Eq.~\ref{eq:gate}):
\begin{equation}
w_{i}^{l,k,k'} =
\frac{\mathbf{g}_i^{l,k} \cdot \mathbf{g}_i^{l,k'}}{\mathbf{g}_i^{l,k}+\mathbf{g}_i^{l,k'}+\epsilon}.
\end{equation}
This keeps the weight small whenever either lens is weakly activated, focusing alignment only on lenses that meaningfully contribute to item $i$'s representation.
As a result, $\mathcal{L}_{\text{CCL}}$ aligns lens outputs across the distinct item subsets (Eq.~\ref{eq:rsb}) without collapsing their individual focus.

\subsubsection{\textbf{Overall Objectives}}

We optimize the SR task using the standard cross-entropy loss $\mathcal{L}_{\text{rec}}$:
\begin{equation}
\mathcal{L}_{\text{rec}} = - \mathbf{y}_u \cdot \log(\hat{\mathbf{y}}_u), \quad \mathcal{L}_{\text{aux}} = \mathcal{L}_{\text{SPCL}} + \mathcal{L}_{\text{CCL}}
\end{equation}
where $\mathbf{y}_u \in \{0,1\}^{|\mathcal{V}|}$ denotes the one-hot ground-truth vector.

The final objective jointly optimizes this main loss with the auxiliary objectives, weighted by a hyperparameter $\lambda$:
\begin{equation}
\mathcal{L} = \mathcal{L}_{\text{rec}} + \lambda \mathcal{L}_{\text{aux}}.
\end{equation}

\begin{table}[h]
\centering
\caption{Statistics of various datasets used in the experiments.}
\renewcommand{\arraystretch}{0.85}
\resizebox{\columnwidth}{!}{
\begin{tabular}{c|ccccc}
\toprule
\textbf{Dataset} & \textbf{\#Users} & \textbf{\#Items} & \textbf{\#Interactions} & \textbf{Avg. Length} & \textbf{Sparsity} \\
\midrule
Toys  & 19,412 & 11,924 & 167,597 & 8.6   & 99.93\% \\
Beauty  & 22,363 & 12,101 & 198,502 & 8.9   & 99.93\% \\
Games   & 24,303 & 10,672 & 231,780  & 9.5   & 99.91\% \\
Sports  & 35,598 & 18,357 & 296,337  & 8.3   & 99.95\% \\
ML-1M   & 6,040  & 3,416  & 999,610  & 165.5 & 95.16\% \\
Yelp  & 29,915  & 21,572  & 345,186  & 11.5 & 99.95\% \\
Electronics  & 192,403 & 63,001 & 1,689,188 & 8.8   & 99.99\% \\
\bottomrule
\end{tabular}
}
\label{tab:dataset_stats}
\end{table}
\section{Experiments}

\subsection{Experimental Setup}

\subsubsection{\textbf{Datasets}}
To evaluate the performance of our proposed model, we conduct experiments on seven widely used benchmark datasets for SR.
The standard benchmarks include five Amazon subcategories~\cite{amazon} (\textit{Toys}, \textit{Beauty}, \textit{Games}, and \textit{Sports} as medium-scale datasets, and \textit{Electronics} as a large-scale industrial dataset), as well as \textit{MovieLens-1M}~\cite{ml-1m} (ML-1M) and \textit{Yelp}~\cite{asghar2016yelp} (interactions after January 1, 2020) to evaluate robustness across platforms.
To mitigate data sparsity issues, we follow~\cite{sasrec,duorec,5core1,5core2, flame} the strategy that retains only users and items with at least five interactions.
Dataset statistics are shown in Tab.~\ref{tab:dataset_stats}.

\definecolor{headercolor}{gray}{0.94}
\definecolor{bestcolor}{RGB}{221,235,247}
\definecolor{secondcolor}{RGB}{252,228,214}

\newcommand{\second}[1]{\underline{#1}}

\newcommand{\best}[1]{\cellcolor{bestcolor}\textbf{#1}}
\begin{table*}[t]
\centering
\caption{Performance comparison of different methods on top-$k$ recommendation. The symbol * denotes statistical significance with p-values $< 0.05$ based on paired t-tests over 5 random seeds.}
\label{tab:performance_comparison}

\small
\setlength{\tabcolsep}{3.2pt}
\renewcommand{\arraystretch}{0.835}

\resizebox{\textwidth}{!}{
\begin{tabular}{cc|ccc|c|cc|cccc|ccc|cc}
\toprule
\rowcolor{headercolor}
\textbf{Dataset} & \textbf{Metric} & \textbf{SASRec} & \textbf{BERT4Rec} & \textbf{GRU4Rec} & \textbf{AC-SASRec} & \textbf{CL4SRec} & \textbf{DuoRec} & \textbf{ICLRec} & \textbf{IOCRec} & \textbf{ICSRec} & \textbf{ELCRec} & \textbf{FAME} & \textbf{FamouSRec} & \textbf{STAR-Rec} & \cellcolor{headercolor!70}\textbf{PRISM} & \cellcolor{headercolor!70}\textit{\textbf{Improv.}} \\
\midrule
\multirow{6}{*}{\textbf{Toys}}
& H@5   & 0.0689 & 0.0487 & 0.0358 & 0.0710 & 0.0732 & 0.0727 & 0.0716 & 0.0726 & 0.0726 & 0.0721 & 0.0734 & 0.0755 & \second{0.0757} & \best{0.0802*} & 5.94\% \\
& H@10  & 0.0911 & 0.0722 & 0.0547 & 0.0953 & 0.0980 & 0.0989 & 0.0985 & 0.0976 & 0.0996 & 0.0965 & 0.0991 & \second{0.1027} & 0.1018 & \best{0.1063*} & 3.51\% \\
& H@20  & 0.1198 & 0.1048 & 0.0809 & 0.1256 & 0.1293 & 0.1306 & 0.1306 & 0.1273 & 0.1305 & 0.1287 & 0.1329 & \second{0.1352} & 0.1345 & \best{0.1410*} & 4.29\% \\
& N@5   & 0.0503 & 0.0333 & 0.0251 & 0.0521 & 0.0528 & 0.0541 & 0.0525 & 0.0524 & 0.0532 & 0.0532 & 0.0526 & 0.0551 & \second{0.0553} & \best{0.0585*} & 5.79\% \\
& N@10  & 0.0577 & 0.0408 & 0.0308 & 0.0602 & 0.0612 & 0.0623 & 0.0616 & 0.0606 & 0.0622 & 0.0605 & 0.0631 & \second{0.0637} & 0.0632 & \best{0.0666*} & 4.55\% \\
& N@20  & 0.0653 & 0.0489 & 0.0377 & 0.0674 & 0.0691 & 0.0700 & 0.0696 & 0.0678 & 0.0700 & 0.0682 & 0.0705 & \second{0.0721} & 0.0717 & \best{0.0751*} & 4.16\% \\
\midrule
\multirow{6}{*}{\textbf{Beauty}}
& H@5   & 0.0597 & 0.0474 & 0.0463 & 0.0634 & 0.0644 & 0.0631 & 0.0642 & 0.0628 & 0.0653 & 0.0646 & 0.0658 & \second{0.0666} & 0.0661 & \best{0.0713*} & 7.06\% \\
& H@10  & 0.0826 & 0.0706 & 0.0692 & 0.0884 & 0.0910 & 0.0910 & 0.0912 & 0.0895 & 0.0918 & 0.0903 & 0.0942 & \second{0.0946} & 0.0943 & \best{0.0996*} & 5.29\% \\
& H@20  & 0.1143 & 0.1008 & 0.0994 & 0.1204 & 0.1251 & 0.1250 & 0.1250 & 0.1228 & 0.1263 & 0.1232 & 0.1288 & 0.1291 & \second{0.1298} & \best{0.1366*} & 5.24\% \\
& N@5   & 0.0421 & 0.0307 & 0.0302 & 0.0450 & 0.0459 & 0.0452 & 0.0464 & 0.0443 & 0.0464 & 0.0460 & 0.0465 & 0.0468 & \second{0.0471} & \best{0.0509*} & 8.07\% \\
& N@10  & 0.0500 & 0.0383 & 0.0377 & 0.0532 & 0.0545 & 0.0541 & 0.0543 & 0.0529 & 0.0546 & 0.0547 & 0.0551 & \second{0.0558} & 0.0553 & \best{0.0603*} & 8.06\% \\
& N@20  & 0.0574 & 0.0459 & 0.0453 & 0.0613 & 0.0630 & 0.0625 & 0.0629 & 0.0612 & 0.0632 & 0.0627 & 0.0637 & \second{0.0647} & 0.0640 & \best{0.0696*} & 7.57\% \\
\midrule
\multirow{6}{*}{\textbf{Games}}
& H@5   & 0.0802 & 0.0722 & 0.0674 & 0.0842 & 0.0872 & 0.0883 & 0.0883 & 0.0879 & 0.0893 & 0.0891 & 0.0902 & 0.0907 & \second{0.0912} & \best{0.0992*} & 8.77\% \\
& H@10  & 0.1211 & 0.1161 & 0.1083 & 0.1248 & 0.1341 & 0.1345 & 0.1348 & 0.1348 & 0.1361 & 0.1352 & 0.1373 & \second{0.1385} & 0.1375 & \best{0.1487*} & 7.36\% \\
& H@20  & 0.1758 & 0.1762 & 0.1643 & 0.1828 & 0.1919 & 0.1966 & 0.1960 & 0.1944 & 0.1969 & 0.1966 & 0.1988 & \second{0.2003} & \second{0.2003} & \best{0.2136*} & 6.64\% \\
& N@5   & 0.0531 & 0.0473 & 0.0441 & 0.0561 & 0.0582 & 0.0592 & 0.0586 & 0.0590 & 0.0593 & 0.0588 & 0.0597 & 0.0602 & \second{0.0607} & \best{0.0659*} & 8.57\% \\
& N@10  & 0.0663 & 0.0614 & 0.0571 & 0.0689 & 0.0732 & 0.0735 & 0.0736 & 0.0737 & 0.0743 & 0.0735 & 0.0749 & 0.0751 & \second{0.0756} & \best{0.0827*} & 9.39\% \\
& N@20  & 0.0803 & 0.0763 & 0.0714 & 0.0834 & 0.0874 & 0.0894 & 0.0887 & 0.0882 & 0.0897 & 0.0899 & 0.0916 & 0.0899 & \second{0.0925} & \best{0.0982*} & 6.16\% \\
\midrule
\multirow{6}{*}{\textbf{Sports}}
& H@5   & 0.0329 & 0.0242 & 0.0220 & 0.0343 & 0.0342 & 0.0358 & 0.0364 & 0.0336 & 0.0358 & 0.0361 & 0.0377 & \second{0.0385} & 0.0378 & \best{0.0416*} & 8.01\% \\
& H@10  & 0.0465 & 0.0381 & 0.0358 & 0.0486 & 0.0526 & 0.0520 & 0.0532 & 0.0513 & 0.0532 & 0.0530 & 0.0547 & \second{0.0553} & 0.0550 & \best{0.0595*} & 7.59\% \\
& H@20  & 0.0662 & 0.0592 & 0.0553 & 0.0686 & 0.0738 & 0.0740 & 0.0746 & 0.0732 & 0.0757 & 0.0737 & 0.0779 & \second{0.0781} & 0.0773 & \best{0.0836*} & 7.04\% \\
& N@5   & 0.0239 & 0.0156 & 0.0145 & 0.0245 & 0.0252 & 0.0253 & 0.0255 & 0.0228 & 0.0261 & 0.0258 & 0.0264 & \second{0.0268} & 0.0266 & \best{0.0291*} & 8.61\% \\
& N@10  & 0.0271 & 0.0197 & 0.0186 & 0.0283 & 0.0301 & 0.0303 & 0.0307 & 0.0276 & 0.0304 & 0.0298 & 0.0312 & \second{0.0322} & 0.0320 & \best{0.0345*} & 7.14\% \\
& N@20  & 0.0324 & 0.0254 & 0.0239 & 0.0339 & 0.0365 & 0.0364 & 0.0367 & 0.0340 & 0.0366 & 0.0367 & 0.0371 & 0.0377 & \second{0.0379} & \best{0.0406*} & 7.12\% \\
\midrule
\multirow{6}{*}{\textbf{ML-1M}}
& H@5   & 0.1584 & 0.1490 & 0.1577 & 0.1653 & 0.1702 & 0.1727 & 0.1770 & 0.1759 & 0.1771 & 0.1778 & 0.1807 & 0.1797 & \second{0.1839} & \best{0.1987*} & 8.05\% \\
& H@10  & 0.2383 & 0.2268 & 0.2297 & 0.2420 & 0.2542 & 0.2545 & 0.2582 & 0.2556 & 0.2611 & 0.2569 & 0.2618 & 0.2697 & \second{0.2712} & \best{0.2883*} & 6.31\% \\
& H@20  & 0.3218 & 0.3263 & 0.3235 & 0.3325 & 0.3496 & 0.3513 & 0.3558 & 0.3545 & 0.3566 & 0.3545 & 0.3721 & \second{0.3771} & 0.3761 & \best{0.3942*} & 4.53\% \\
& N@5   & 0.1060 & 0.0958 & 0.1020 & 0.1099 & 0.1150 & 0.1182 & 0.1181 & 0.1180 & 0.1189 & 0.1176 & 0.1202 & 0.1213 & \second{0.1253} & \best{0.1354*} & 8.06\% \\
& N@10  & 0.1318 & 0.1211 & 0.1251 & 0.1347 & 0.1418 & 0.1442 & 0.1447 & 0.1435 & 0.1459 & 0.1427 & 0.1478 & 0.1503 & \second{0.1535} & \best{0.1645*} & 7.17\% \\
& N@20  & 0.1530 & 0.1456 & 0.1488 & 0.1572 & 0.1662 & 0.1685 & 0.1670 & 0.1701 & 0.1675 & 0.1674 & 0.1754 & 0.1767 & \second{0.1772} & \best{0.1913*} & 7.96\% \\
\midrule
\multirow{6}{*}{\textbf{Yelp}}
& H@5   & 0.0275 & 0.0263 & 0.0250 & 0.0281 & 0.0294 & 0.0296 & 0.0302 & 0.0312 & 0.0317 & 0.0289 & 0.0328 & 0.0335 & \second{0.0339} & \best{0.0371*} & 9.44\% \\
& H@10  & 0.0453 & 0.0459 & 0.0435 & 0.0475 & 0.0509 & 0.0519 & 0.0502 & 0.0542 & 0.0543 & 0.0494 & 0.0570 & 0.0576 & \second{0.0581} & \best{0.0636*} & 9.47\% \\
& H@20  & 0.0744 & 0.0766 & 0.0720 & 0.0771 & 0.0855 & 0.0865 & 0.0835 & 0.0889 & 0.0892 & 0.0878 & 0.0942 & \second{0.0949} & 0.0947 & \best{0.1041*} & 9.69\% \\
& N@5   & 0.0170 & 0.0163 & 0.0156 & 0.0173 & 0.0181 & 0.0187 & 0.0186 & 0.0191 & 0.0199 & 0.0181 & 0.0207 & \second{0.0210} & 0.0204 & \best{0.0232*} & 10.48\% \\
& N@10  & 0.0230 & 0.0230 & 0.0214 & 0.0237 & 0.0253 & 0.0253 & 0.0253 & 0.0262 & 0.0268 & 0.0263 & 0.0276 & \second{0.0287} & 0.0280 & \best{0.0319*} & 11.15\% \\
& N@20  & 0.0301 & 0.0304 & 0.0285 & 0.0312 & 0.0336 & 0.0342 & 0.0330 & 0.0360 & 0.0360 & 0.0344 & 0.0361 & 0.0372 & \second{0.0376} & \best{0.0418*} & 11.17\% \\
\midrule
\multirow{6}{*}{\textbf{Elec.}}
& H@5   & 0.0312 & 0.0254 & 0.0279 & 0.0338 & 0.0352 & 0.0353 & 0.0350 & 0.0349 & 0.0357 & 0.0353 & 0.0362 & \second{0.0374} & 0.0373 & \best{0.0398*} & 6.42\% \\
& H@10  & 0.0453 & 0.0389 & 0.0414 & 0.0485 & 0.0508 & 0.0504 & 0.0502 & 0.0508 & 0.0510 & 0.0499 & 0.0529 & \second{0.0539} & 0.0536 & \best{0.0556*} & 3.15\% \\
& H@20  & 0.0634 & 0.0572 & 0.0605 & 0.0685 & 0.0708 & 0.0714 & 0.0704 & 0.0717 & 0.0707 & 0.0702 & 0.0742 & 0.0755 & \second{0.0756} & \best{0.0779*} & 3.04\% \\
& N@5   & 0.0216 & 0.0168 & 0.0188 & 0.0237 & 0.0245 & 0.0246 & 0.0244 & 0.0238 & 0.0248 & 0.0247 & 0.0248 & \second{0.0259} & 0.0256 & \best{0.0272*} & 5.02\% \\
& N@10  & 0.0262 & 0.0212 & 0.0232 & 0.0284 & 0.0294 & 0.0293 & 0.0293 & 0.0290 & 0.0297 & 0.0294 & 0.0304 & \second{0.0313} & 0.0312 & \best{0.0323*} & 3.19\% \\
& N@20  & 0.0324 & 0.0257 & 0.0280 & 0.0340 & 0.0345 & 0.0347 & 0.0343 & 0.0345 & 0.0347 & 0.0345 & 0.0360 & \second{0.0368} & 0.0367 & \best{0.0383*} & 4.08\% \\
\bottomrule
\end{tabular}}
\end{table*}

\subsubsection{\textbf{Evaluation Settings}}
We follow the leave-one-out evaluation protocol~\cite{sasrec,loo1,loo2}: for each user, the last item is held out for testing, the second-to-last for validation, and the rest for training.
We evaluate with Hit Rate (H@$K$) and Normalized Discounted Cumulative Gain (N@$K$), reporting averages over 5 random seeds for $K\in\{5,10,20\}$ without negative sampling~\cite{negativesampling}.

\subsubsection{\textbf{Baselines}}
To verify the effectiveness of our proposed model, \modelname, we compare it against a comprehensive set of state-of-the-art baseline models from five distinct categories:
\begin{itemize}[leftmargin=*]
    \item \textbf{General SR}: This category includes foundational and widely recognized models for SR. We include GRU4Rec~\cite{gru4rec}, an RNN-based approach, as well as dominant transformer-based models like SASRec~\cite{sasrec} and BERT4Rec~\cite{bert4rec}.
    \item \textbf{Attention Calibrated SR}: We compare against AC-SASRec~\cite{actsr}, which addresses the issue of misaligned attention distributions by introducing a calibration to refine the self-attention mechanism.
    \item \textbf{Contrastive Learning based SR}: This category features models that employ contrastive learning for robust representation learning, including CL4SRec~\cite{cl4srec} and DuoRec~\cite{duorec}.
    \item \textbf{Intent-based SR}: We compare against recent models that explicitly model the intent prototypes to consider the user preferences, such as ICLRec~\cite{iclrec}, IOCRec~\cite{iocrec}, ICSRec~\cite{icsrec}, and ELCRec~\cite{elcrec}.
    \item \textbf{MoE-based SR}: To compare against models with similar architectural principles and capacity, we include MoE-based recommendation models such as FAME~\cite{fame}, FamouSRec~\cite{famousrec}, and STAR-Rec~\cite{starrec}.
\end{itemize}

\subsubsection{\textbf{Implementation Details}}
All models are implemented in PyTorch and trained with the Adam optimizer~\cite{adam}, using a learning rate of 0.001, $\beta_1$=0.9, $\beta_2$=0.999, a batch size of 256, a dropout rate of 0.5, and a patience of 10 epochs for early stopping on N@20.
Following the traditional SR settings~\cite{bert4rec,duorec,cl4srec,icsrec,elcrec}, we adopt $L{=}2$ transformer layers and 2 attention heads, with a hidden dimension of $d$=64, yielding 1 PRISM layer, and the max sequence length is set to 50.
We search for the number of perspective lenses $K \in \{2,4,6,8\}$, the loss weight $\lambda \in \{0.01, 0.05, 0.1, 0.3, 0.5\}$, and the temperature $\tau \in \{0.05, 0.1, 0.7, 1.0, 2.0\}$.
The perspective dropout rate $\rho$ is set to 0.25, and we use in-batch target items as $\mathbf{E}_{\text{sample}}$ for $\mathcal{L}_{\text{CCL}}$.
Other hyperparameters for all baseline models are tuned according to the search spaces provided in their original papers.
All experiments are conducted on a single RTX 3090.

\subsection{Performance Comparison}

PRISM consistently outperforms all baselines across datasets and metrics (Tab.~\ref{tab:performance_comparison}).
Detailed analysis follows:

General SR models such as SASRec, BERT4Rec, and GRU4Rec perform well but struggle with diverse inter-item relations.
AC-SASRec calibrates attention distributions with a learnable matrix but lacks explicit relational guidance, treating the symptom rather than the structural cause.
CL4SRec and DuoRec achieve gains via sequence-level contrastive learning, but their sequence-level augmentation misses item-level relational structure.

Intent-based SR models perform competitively, confirming the value of modeling user preference beyond raw item sequences.
However, they rely on fixed prototypes and sequence-level modeling, overlooking fine-grained inter-item relations underlying diverse preferences.
ICSRec seeks to capture coarse and fine-grained patterns, but its prototypes cannot represent nuanced preferences embedded in item-level relations.
PRISM addresses this through item-level Dual-View analysis, refining homogeneous relations and recovering overlooked heterogeneous ones, while $\mathcal{L}_{\text{SPCL}}$ provides sequence-level preference alignment.

The strong performance of MoE baselines like FAME, FamouSRec, and STAR-Rec confirms that expanding model capacity benefits the modeling of diverse user behaviors. 
Yet capacity alone is not the determining factor: PRISM achieves the strongest results with fewer parameters than these baselines (more detailed analysis in Sec.~\ref{sec:scalability}), demonstrating that how the added capacity is used matters more than its size, and that directing it toward inter-item relational modeling is effective.

\subsection{Further Analysis}

\begin{table}[t]
  \centering
  \caption{Ablation study of PRISM training objectives on Toys and ML-1M datasets.}
  \label{tab:ablation_study_loss}
    \renewcommand{\arraystretch}{0.8}
\setlength{\tabcolsep}{3.0pt}
  \resizebox{\columnwidth}{!}{
  \begin{tabular}{lcccc}
    \toprule
    \multirow{2.5}{*}{\textbf{Models}} & \multicolumn{2}{c}{\textbf{Toys}} & \multicolumn{2}{c}{\textbf{ML-1M}} \\
    \cmidrule{2-3} \cmidrule{4-5}
    & H@20 & N@20 & H@20 & N@20 \\
    \midrule
    PRISM & \textbf{0.1410*} & \textbf{0.0751*} & \textbf{0.3942*} & \textbf{0.1913*} \\
    \midrule
    \textit{w/o} $\mathcal{L}_\text{SPCL}$ & 0.1362 & 0.0732 & 0.3776 & 0.1804 \\
    $\rightarrow$\textit{w/} $\mathcal{L}_\text{SPCL}$ \textit{w/o} Same Target & 0.1401 & 0.0746 & 0.3829 & 0.1833 \\
    $\rightarrow$\textit{w/} $\mathcal{L}_\text{SPCL}$ \textit{w/o} SPM & 0.1395 & 0.0740 & 0.3844 & 0.1852 \\
    \textit{w/o} $\mathcal{L}_\text{CCL}$ & 0.1403 & 0.0744 & 0.3887 & 0.1878 \\
    \bottomrule
  \end{tabular}
  }
\end{table}\begin{figure}[t]
    \centering
    \includegraphics[width=\columnwidth]{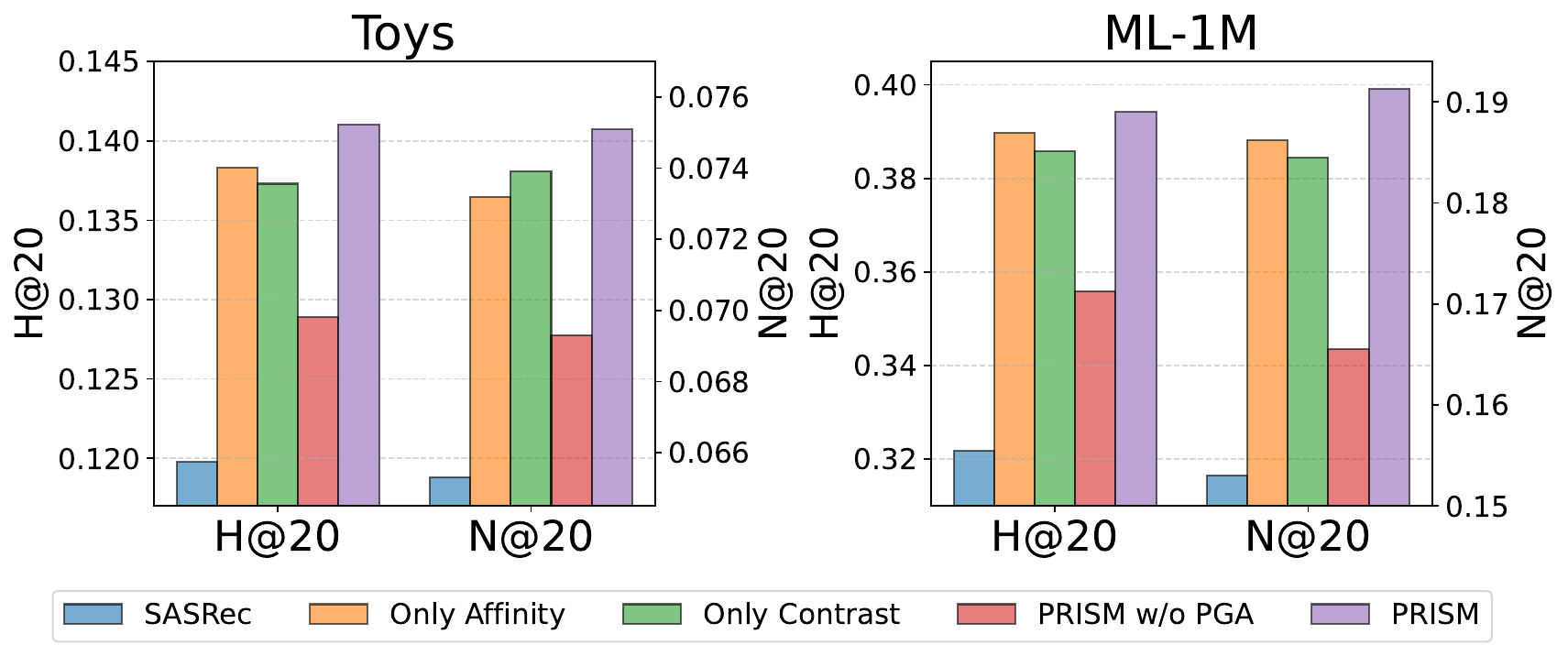}
    \caption{Ablation study of PRISM's dual view design on the Toys and ML-1M datasets.}
    \label{fig:view_ablation}
\end{figure}

\subsubsection{\textbf{Revisiting the Preliminary Diagnostics}}
We revisit the diagnostics in Sec.~\ref{sec:analysis} to verify whether PRISM alleviates the observed attention failures.
Fig.~\ref{fig:recovery} shows that the attention interventions (C1--C4) yield near-zero or even negative $\Delta_{\text{confidence}}$ for PRISM.
Unlike the baselines, where any redistribution of attention helped prediction, perturbing PRISM's attention provides no benefit and can even degrade it.
This indicates that PRISM already attends to the items that are causally important for target prediction; disturbing these well-calibrated relations removes useful evidence rather than adding missing signals.
Fig.~\ref{fig:alignment} further corroborates this: PRISM improves both $\rho(A,I)$ and Align@1 over all baselines, confirming that its attention distribution is now aligned with the items the model actually relies on for prediction.

\begin{table*}[t]
\centering
\caption{Ranking performance and sensitivity analysis on real users and item IDs (ML-1M). We analyze the model's confidence drop ($\Delta \text{Conf}$) when masking the \textbf{\textit{Bridge Item}} (marked in \textcolor{red}{\textbf{Red}}) that contains the bridging \textcolor{blue}{\textbf{minor genre}} required for the target.}
\label{tab:case_study}
\resizebox{\textwidth}{!}{
\begin{tabular}{c|l|c|cc|cc|cc}
\toprule
\multirow{2}{*}{\textbf{User}} & \multicolumn{1}{c|}{\multirow{2}{*}{\textbf{Interaction Sequence (Item ID$_{\text{GenreID}}$)}}} & \multirow{2}{*}{\textbf{Target}} & \multicolumn{2}{c|}{\textbf{SASRec}} & \multicolumn{2}{c|}{\textbf{ICSRec}} & \multicolumn{2}{c}{\textbf{PRISM (Ours)}} \\
 & & & Rank & $\Delta$Conf & Rank & $\Delta$Conf & Rank & $\Delta$Conf \\
\midrule
\multirow{2}{*}{5979} & $1794_{\text{(G1,G2)}} \to 1959_{\text{(G2,G3)}} \to 345_{\text{(G1,G2)}} \to 410_{\text{(G1)}} \to 1081_{\text{(G1)}} \to$ & \multirow{2}{*}{$3033_{\text{(G1,\textcolor{blue}{\textbf{G6}})}}$} & \multirow{2}{*}{435} & \multirow{2}{*}{-25.42\%} & \multirow{2}{*}{327} & \multirow{2}{*}{-3.26\%} & \multirow{2}{*}{\textbf{3}} & \multirow{2}{*}{\textbf{-43.62\%}} \\
 & $2362_{\text{(G2)}} \to \textcolor{red}{\textbf{1924}}_{\text{(G5,\textcolor{blue}{\textbf{G6}})}} \to 319_{\text{(G7)}} \to 2712_{\text{(G2)}} \to 597_{\text{(G1,G3)}}$ & & & & & & & \\
\midrule
\multirow{2}{*}{1397} & $\textcolor{red}{\textbf{237}}_{\text{(G8,\textcolor{blue}{\textbf{G9}})}} \to 203_{\text{(G1,G3)}} \to 574_{\text{(G8)}} \to 1306_{\text{(G5,G8)}} \to 92_{\text{(G1)}} \to$ & \multirow{2}{*}{$386_{\text{(G1,\textcolor{blue}{\textbf{G9}})}}$} & \multirow{2}{*}{197} & \multirow{2}{*}{-10.57\%} & \multirow{2}{*}{22} & \multirow{2}{*}{-5.66\%} & \multirow{2}{*}{\textbf{9}} & \multirow{2}{*}{\textbf{-12.17\%}} \\
 & $34_{\text{(G1)}} \to 1611_{\text{(G1)}} \to 290_{\text{(G1)}} \to 521_{\text{(G1,G2)}} \to 626_{\text{(G1,G2)}}$ & & & & & & & \\
\bottomrule
\end{tabular}
}
\end{table*}

\subsubsection{\textbf{Ablation of Training Objectives}}

To verify the role of each training objective, we ablate $\mathcal{L}_{\text{SPCL}}$ and $\mathcal{L}_{\text{CCL}}$ individually (Tab.~\ref{tab:ablation_study_loss}).
Removing $\mathcal{L}_{\text{SPCL}}$ entirely causes the largest performance drop, confirming that aligning users who share the same target is critical for capturing sequence-level preference.
Within $\mathcal{L}_{\text{SPCL}}$, its two sub-components serve complementary roles.
Without \textit{Same Target} augmentation, performance degrades, showing that users with a shared target effectively provide each other's sequence-level preference signal.
Without \textit{Stochastic Perspective Masking} (SPM), the contrastive loss over-aligns all perspectives into a single shared pattern, collapsing the distinct relational patterns each lens captures.
As a result, users who arrive at the same target through different item compositions become indistinguishable.
Removing $\mathcal{L}_{\text{CCL}}$ also lowers performance, confirming that without distribution alignment, the weighted sum of lens outputs mixes incompatible representation spaces, degrading the final item representation.

\subsubsection{\textbf{Ablation of Dual-View Analysis}}

To evaluate the individual contributions of the \textit{Dual-View Analysis}, we design two variants (Fig.~\ref{fig:view_ablation}): \textit{Only Affinity}, which is forced to use only the primary (PSA) lens of the items by setting all other routing logits to $-\infty$ (Eq.~\ref{eq:gate}), and \textit{Only Contrast}, which retains only non-PSA lenses.
Remarkably, while even single-view variants outperform the baseline, integrating both yields the highest performance.
Each variant improves over the baseline through a distinct mechanism: \textit{Only Affinity} refines the homogeneous relations already captured by SASRec, while \textit{Only Contrast} recovers the heterogeneous relations that SASRec overlooks.
To isolate the contribution of the calibration itself from increased capacity, we further test \textit{w/o PGA}, which removes the relational boost signal $\mathbf{S}^{l,k}$ (Eq.~\ref{eq:pga}) while keeping the parameter count identical to the full PRISM. 
\textit{w/o PGA} performs worse than even the single-view variants on both datasets, confirming that PRISM's gains originate from the perspective-guided calibration of attention, the mechanism that directly addresses the overlooked relations identified in Sec.~\ref{sec:analysis}.

\subsubsection{\textbf{Visualizing the Semantic Anchor Router}}

\begin{figure}[t]
    \centering
    \begin{subfigure}{0.48\columnwidth}
        \centering
        \includegraphics[width=0.9\linewidth]{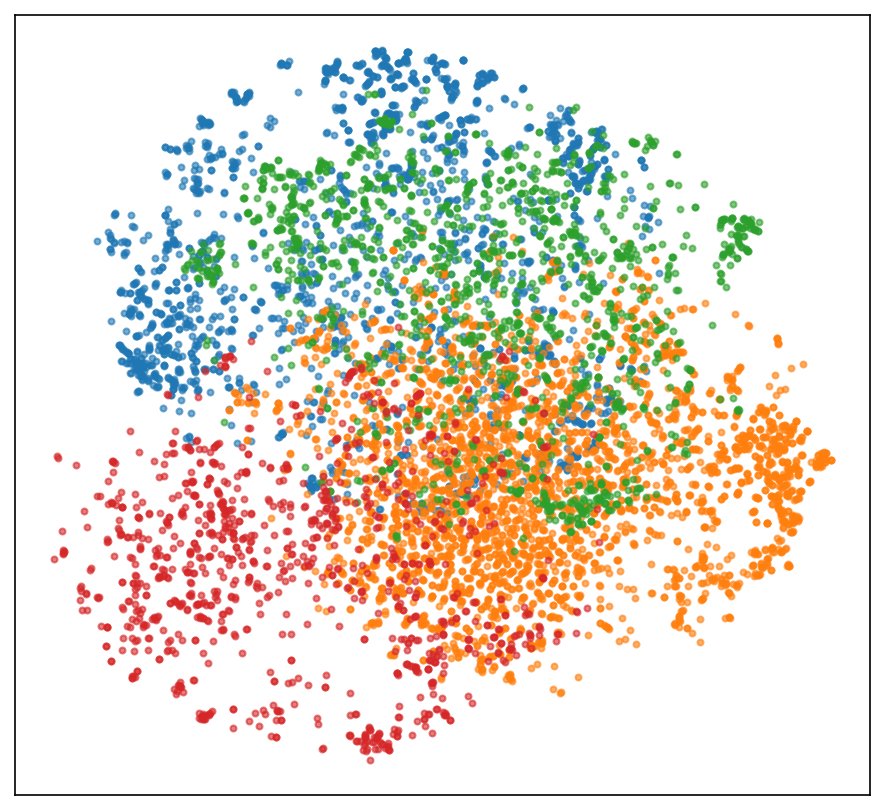}
        \caption{Toys}
    \end{subfigure}
    \hfill
    \begin{subfigure}{0.48\columnwidth}
        \centering
        \includegraphics[width=0.9\linewidth]{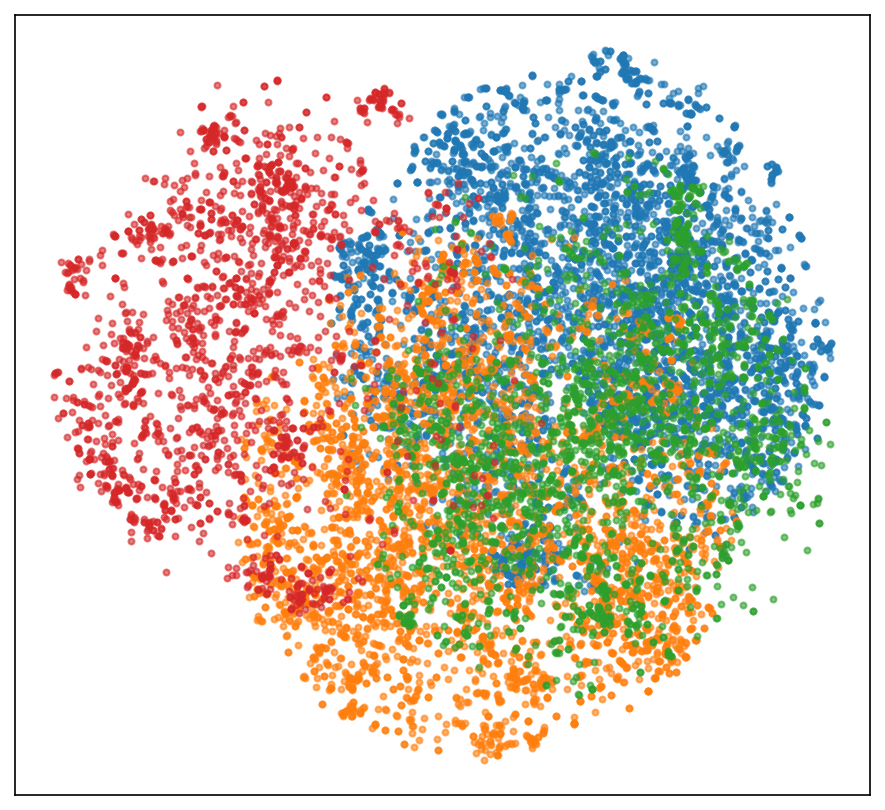}
        \caption{Sports}
    \end{subfigure}
    \caption{t-SNE projection of item embeddings colored by PSA assignment ($K=4$).}
    \label{fig:tsne}
\end{figure}

To examine whether the Semantic Anchor Router (Sec.~\ref{sec:sar}) learns meaningful groupings, we visualize the item embeddings via t-SNE on Toys and Sports, coloring each item by its PSA assignment (Fig.~\ref{fig:tsne}).
Although PRISM is not designed as a clustering method, the lens groups form clearly separated regions, showing that the router consistently assigns items into distinct semantic groups.
Moreover, items are spread evenly across the four groups, as the noisy gating mechanism (Eq.~\ref{eq:logits}) promotes balanced exploration across lenses during training.
This balanced routing structure enables the Dual-View synergy to operate effectively: clear PSA boundaries give the \textit{Affinity View} coherent within-group material to refine, while ensuring the \textit{Contrast View} has sufficient cross-group diversity.

\begin{table}[t]
  \centering
  \caption{Performance on Conventional (top 50\%, high sequence-target similarity) and Novel (bottom 20\%, low similarity) ML-1M subsets, testing robustness to target preferences not explicit in the sequence.}
  \label{tab:conventional}
  \resizebox{\columnwidth}{!}{
  \begin{tabular}{lccccc}
    \toprule
    \multirow{2.4}{*}{\textbf{Models}} & \multicolumn{2}{c}{\textbf{Conventional Set}} & \multicolumn{2}{c}{\textbf{Novel Set}} & \multirow{2.4}{*}{\textbf{All (Orig.)}} \\
    \cmidrule(lr){2-3} \cmidrule(lr){4-5}
    & N@20 & vs. All (\%) & N@20 & vs. All (\%) \\
    \midrule
    SASRec & 0.1098 & \textit{-28.24} & 0.0572 & \textit{-62.61} & 0.1530\\
    ICSRec & 0.1766 & \textit{5.43} & 0.0484 & \textit{-71.1} & 0.1675\\
    PRISM & \textbf{0.2028*} & \textit{6.01} & \textbf{0.0797*} & \textit{-58.34} & \textbf{0.1913*}\\
    \bottomrule
  \end{tabular}
  }
\end{table}

\subsubsection{\textbf{Analysis of Mitigating Similarity Bias}}

To understand the impact of this mitigation on recommendation outcomes, we present a qualitative and quantitative analysis.

\subparagraph{\textbf{Qualitative Case Study.}}
To examine this effect using real-user examples, we conduct a qualitative case study on ML-1M (Tab.~\ref{tab:case_study}).
In these sequences, the target shares a sparse minor genre (e.g., G6, G9) with a multi-category \textit{Bridge Item} that is semantically distant from the surrounding dominant-genre items (e.g., G1, G2); predicting the target thus hinges on recovering the heterogeneous relation that similarity bias suppresses.
SASRec's attention concentrates on dominant-genre items similar to each other, marginalizing the \textit{Bridge Item}, while ICSRec's sequence-level preference is similarly dominated by these genres. Both rank the minor-genre target poorly and show little confidence drop when the \textit{Bridge Item} is masked, confirming the signal had been discarded.
PRISM, in contrast, ranks the targets highly and shows a substantial drop when the \textit{Bridge Item} is masked, treating it as causal evidence.
Its Perspective Lenses decompose the \textit{Bridge Item} into its multiple semantic groups, surfacing the minor-genre signals (G6, G9) suppressed by similarity bias; the \textit{Contrast View} then synthesizes these with the dominant-genre context of surrounding items, enabling prediction of targets that span both groups (G1+G6, G1+G9).

\subparagraph{\textbf{Quantitative Analysis.}}
In Tab.~\ref{tab:conventional}, we partition the test set by sequence-target similarity:
a \textit{Conventional Set} (top 50\%, high similarity), where the target aligns with the user's surface-level preference and overall sequence modeling suffices for prediction, and a \textit{Novel Set} (bottom 20\%, low similarity), where the target reflects preferences not explicit in the sequence and requires going beyond surface signals.
On the \textit{Conventional Set}, sequence-level intent-based ICSRec benefits from global modeling, but PRISM achieves the highest performance by combining sequence-level preference from $\mathcal{L}_{\text{SPCL}}$ with homogeneous refinement via the \textit{Affinity View}.
On the \textit{Novel Set}, where targets require the kind of cross-group recovery the case study illustrates, ICSRec degrades more than even SASRec, as its global intent actively suppresses the non-dominant signals these targets depend on.
PRISM remains the most robust through its \textit{Contrast View} complemented by $\mathcal{L}_{\text{SPCL}}$.

\subsubsection{\textbf{Analysis Across Sequence Lengths}}

\begin{figure}[t]
    \centering
    \includegraphics[width=\columnwidth]{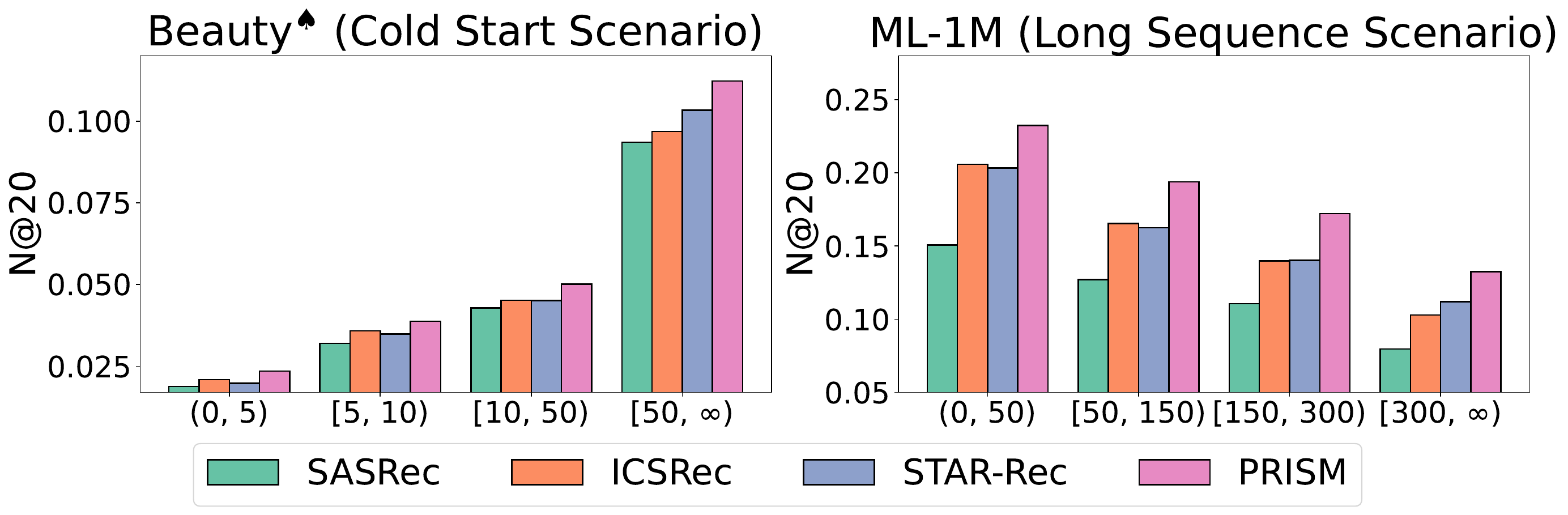}
    \caption{Analysis of performance in cold start and long sequence scenarios on Beauty$^\spadesuit$ and ML-1M datasets.}
    \label{fig:seq_len}
\end{figure}

We evaluate PRISM's robustness across two challenging regimes of sequence length on Beauty$^\spadesuit$\footnote{The Beauty$^\spadesuit$ dataset is processed using a single 5-core filtering pass, commonly used in cold-start settings~\cite{cold1,cold2}, unlike the iteratively filtered version used in Tab.~\ref{tab:dataset_stats}.} and ML-1M (Fig.~\ref{fig:seq_len}). These regimes pose complementary challenges: sparse evidence in short sequences and increasingly diverse preferences in long sequences.

\subparagraph{\textbf{Cold Start Scenario.}}
In the cold start scenario, where each user has only a handful of past interactions, PRISM consistently outperforms all baselines across the entire short-sequence range.
The advantage is most pronounced in the extreme regime ($T<5$), where PRISM improves over ICSRec by 12.44\%, demonstrating its effectiveness with sparse evidence.
With so few items, even a single cross-group relation can be decisive for prediction; the Dual-View Analysis ensures that both homogeneous and heterogeneous relations are extracted from every available pair, recovering preference from signals that a single attention pass would overlook.

\subparagraph{\textbf{Long Sequence Scenario.}}
In the long sequence scenario, PRISM again surpasses all baselines in every setting.
As sequence length increases from $T<50$ to $T\ge300$, ICSRec suffers the largest drop ($-50.02\%$), confirming that a fixed set of pre-defined intent prototypes lacks the capacity to represent the diverse preferences that accumulate over many interactions.
PRISM, in contrast, yields the smallest degradation ($-42.99\%$): it synthesizes user preference dynamically from item relations at each step, and its expressivity is bounded not by $K$ itself but by the combinatorial space of how lenses combine across the sequence, so representational capacity scales naturally with length.
These results confirm PRISM's robustness across both extremes of sequence length.

\subsubsection{\textbf{Scalability and Efficiency Analysis}}
\label{sec:scalability}
We evaluate PRISM's scalability and efficiency empirically and theoretically when increasing $K$ (Tab.~\ref{tab:scalability}).

\subparagraph{\textbf{Additional Parameters.}}
The PRISM layer introduces three components:
\begin{itemize}[nosep, leftmargin=*]
    \item \textbf{Semantic Anchor Router}: two matrices in $\mathbb{R}^{d \times K}$ ($\mathbf{W}_\mathbf{g}^l$, $\mathbf{W}_{\text{noise}}^l$).
    \item \textbf{Perspective Lenses}: a learnable scalar $\beta^l$ and a linear layer in $\mathbb{R}^{d \times d}$.
    \item \textbf{Feed-Forward Network}: two matrices in $\mathbb{R}^{d \times 4d}$ and $\mathbb{R}^{4d \times d}$.
\end{itemize}
By reusing $\mathbf{W}_v^l$ and $\mathbf{A}^l$ from the previous transformer layer, PRISM increases the parameter count by only $2d$ per lens; increasing $K$ from 2 to 8 adds only 768 parameters in our settings.

\begin{table}[t]
\centering
\caption{Scalability analysis across different $K$ values on the Toys dataset.}
\label{tab:scalability}
\setlength{\tabcolsep}{3.2pt}
\resizebox{\columnwidth}{!}{
\begin{tabular}{lcccccc}
\toprule
\multirow{2}{*}{\textbf{Model}} & \multirow{2}{*}{\textbf{Params}} & \textbf{Train Time} & \textbf{Infer. Time} & \multicolumn{2}{c}{\textbf{Peak VRAM (GB)}} & \textbf{Perf.} \\
 &  & (s/epoch) & (s) & Train & Infer & N@20 \\
\midrule
SASRec & 866,496 & 10.20 & 3.43 & 1.4 & 1.2 & 0.0653 \\
FamouSRec & 964,739 & 103.55 & 4.47 & 4.5 & 1.4 & 0.0721 \\
STAR-Rec & 1,095,948 & 35.12 & 8.26 & 4.5 & 2.5 & 0.0717 \\
\midrule
PRISM ($K=2$) & 908,801 & 18.38 & 4.00 & 2.3 & 1.8 & 0.0731 \\
PRISM ($K=4$) & 909,057 & 20.99 & 4.13 & 3.1 & 2.1 & \textbf{0.0751*} \\
PRISM ($K=6$) & 909,313 & 22.59 & 4.45 & 4.1 & 2.3 & 0.0728 \\
PRISM ($K=8$) & 909,569 & 25.27 & 4.73 & 5.2 & 2.6 & 0.0744 \\
\bottomrule
\end{tabular}
}
\end{table}

\begin{table}[t]
\centering
\caption{Time and space complexity of each component.}
\label{tab:complexity}
\resizebox{\columnwidth}{!}{
\begin{tabular}{lcc}
\toprule
\textbf{Component} & \textbf{Time} & \textbf{Space} \\
\midrule
Transformer layer & $O(T^2d + Td^2)$ & $O((T+d)^2)$ \\
PRISM layer & $O(T^2Kd + TKd^2)$ & $O(K(T+d)^2)$ \\
$\mathcal{L}_{\text{SPCL}}$ & $O(BK(T^2d + Td^2) + B^2d)$ & $O(BTK + B^2)$ \\
$\mathcal{L}_{\text{CCL}}$ & $O(BTKNd)$ & $O(BTKN + BTK^2)$ \\
\bottomrule
\end{tabular}
}
\end{table}

\begin{figure}[t]
    \centering
    \includegraphics[width=\columnwidth]{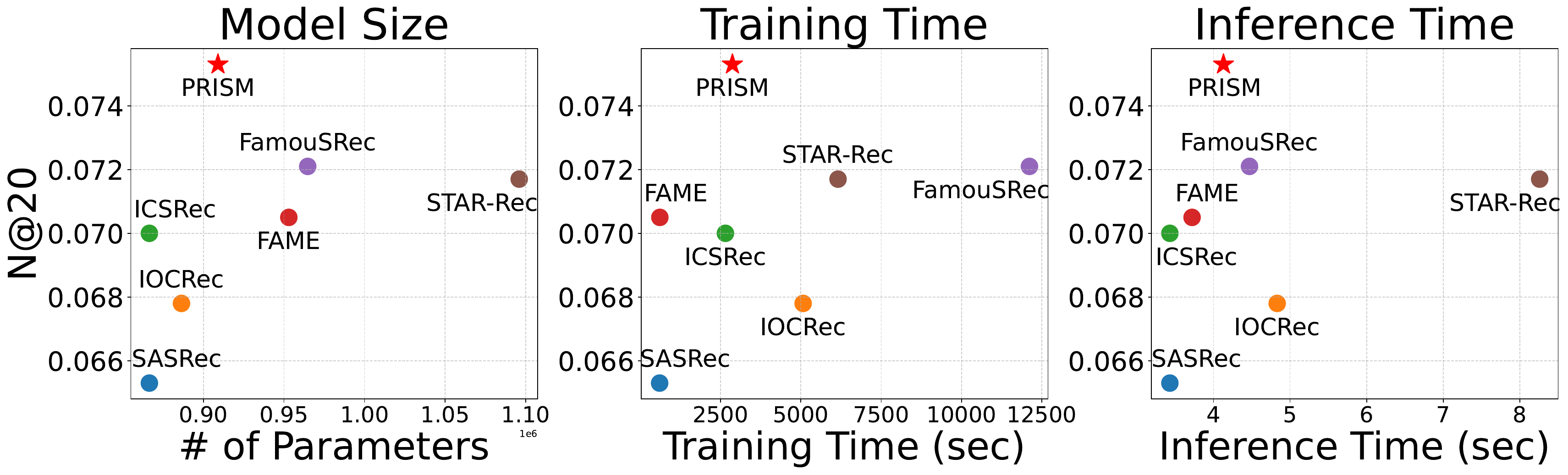}
    \caption{Comparison of performance vs. (a) model size, (b) training time until convergence, and (c) inference time across baseline models and PRISM on the Toys dataset.}
    \label{fig:capacity}
\end{figure}

\subparagraph{\textbf{Theoretical Complexity and Empirical Cost.}}
Tab.~\ref{tab:complexity} reports the time and space complexity of each component; the PRISM layer scales linearly with $K$.
In practice, weight sharing across lenses substantially mitigates this overhead: training time, inference time, and VRAM usage increase with $K$ but remain practical compared with MoE baselines (Tab.~\ref{tab:scalability}).
PRISM surpasses the strongest baseline FamouSRec across all $K$ settings: even the smallest $K{=}2$ already outperforms FamouSRec, and $K{=}4$ achieves the best performance while using fewer parameters (909K vs. 965K).
Moreover, PRISM trains $5\times$ faster than FamouSRec and, at $K{=}4$, uses less training memory than STAR-Rec (3.1 GB vs. 4.5 GB), while maintaining inference latency close to SASRec.
Fig.~\ref{fig:capacity} extends this comparison to non-MoE baselines, where PRISM achieves the highest N@20 at competitive cost across all three dimensions.
$K$ can thus be selected by balancing semantic capacity against computational budget, with overhead remaining acceptable relative to competitive baselines.

\subparagraph{\textbf{Scaling to Industrial Data.}}
On the industrial-scale \textit{Electronics} dataset, clustering-based intent models (e.g., ICLRec, ICSRec) require substantially more prototypes and longer training, as their clustering must cover a far larger pool of users (192K in Electronics vs. 19K in Toys), while MoE-based baselines incur proportionally heavier overhead from their parallel expert architectures.
PRISM achieves the highest accuracy at a lower training cost, confirming that its efficiency advantage widens with data size.
\subsubsection{\textbf{Hyperparameter Sensitivity}}

\begin{figure}[t]
    \centering
    \includegraphics[width=\columnwidth]{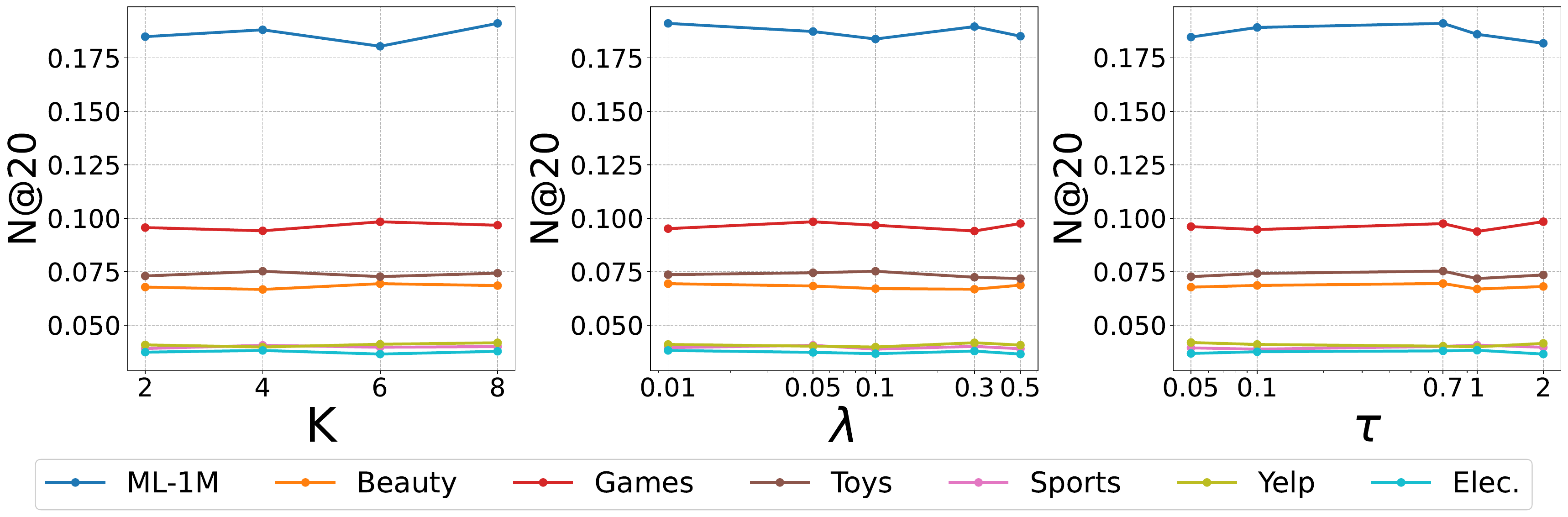}
    \caption{Impact of hyperparameters $K$, $\lambda$, and $\tau$.}
    \label{fig:hyperparameter}
\end{figure}

In Fig.~\ref{fig:hyperparameter}, we conduct a sensitivity analysis for PRISM's robustness across all seven datasets by varying its three hyperparameters: the number of Perspective Lenses $K$, the auxiliary loss weight $\lambda$, and the temperature $\tau$ of $\mathcal{L}_{\text{SPCL}}$.
Performance remains stable across a wide range of values and consistently exceeds the strongest baseline in nearly all configurations, indicating that PRISM is robust to hyperparameter choice and requires minimal tuning effort for diverse domains in practice.

\section{Conclusion}

In this paper, we empirically reveal that transformer-based sequential recommendation models systematically overlook inter-item relations that carry meaningful evidence for target prediction.
To enrich these relations, we propose \textbf{PRISM}, a module placed between transformer layers that complements self-attention through $K$ \textit{Perspective Lenses}. Each lens operates from one of two complementary views: the \textit{Affinity View} refines homogeneous relations among semantically related items, while the \textit{Contrast View} surfaces heterogeneous relations that similarity bias would otherwise suppress. 
With $\mathcal{L}_{\text{SPCL}}$ for sequence-level preference alignment and $\mathcal{L}_{\text{CCL}}$ for cross-lens coherence, PRISM captures user preference from item-level relational signals to sequence-wide behavioral patterns.
Experiments on seven benchmark datasets demonstrate that PRISM outperforms state-of-the-art baselines across various real-world scenarios.
\begin{acks}
This work was supported by the National Research Foundation of Korea (NRF) grant funded by the Korea government (MSIT) (No. RS-2024-00335873).
This work was also supported by Institute of Information \& communications Technology Planning \& Evaluation (IITP) grant funded by the Korea government. (MSIT) (No.RS-2019-II191906, Artificial Intelligence Graduate School Program(POSTECH))
\end{acks}
\section{GenAI Usage Disclosure}

We used ChatGPT and Claude for grammar correction, language polishing, and code refactoring. Ideation, methodology, and experimental design were carried out by the authors without GenAI assistance.

\bibliographystyle{ACM-Reference-Format}
\bibliography{reference}

\end{document}